\documentclass{aa}
\usepackage[T1]{fontenc}
\usepackage[active]{srcltx}
\usepackage{url}
\usepackage{amsmath}
\usepackage{graphicx}

\makeatletter

\usepackage[varg]{txfonts}  
\bibpunct{(}{)}{;}{a}{}{,}  
\usepackage[dvipdfm, breaklinks, colorlinks, citecolor=blue]{hyperref} 

\makeatother

\begin{document}

\title{Spectrum radial velocity analyser (SERVAL)}

\subtitle{High-precision radial velocities and two alternative spectral indicators}

\abstract{The CARMENES survey is a high-precision radial velocity (RV) programme
that aims to detect Earth-like planets orbiting low-mass stars.}{We
develop least-squares fitting algorithms to derive the RVs and additional
spectral diagnostics implemented in the SpEctrum Radial Velocity Analyser (SERVAL), a publicly available python
code.}{We measured the RVs using high signal-to-noise templates
created by coadding all available spectra of each star.  We define the chromatic index
as the RV gradient as a function of wavelength with the
RVs measured in the echelle orders. Additionally, we computed
the differential line width by correlating the fit residuals with
the second derivative of the template to track variations in the stellar
line width.}{Using HARPS data, our SERVAL code achieves a RV precision
at the level of 1\,m/s. Applying the chromatic index to CARMENES
data of the active star YZ~CMi, we identify apparent RV variations
induced by stellar activity. The differential line width is found
to be an alternative indicator to the commonly used full width half
maximum.}{We find that at the red optical wavelengths (700--900\,nm)
obtained by the visual channel of CARMENES, the chromatic index is
an excellent tool to investigate stellar active regions and to identify
and perhaps even correct for activity-induced RV variations.}

\author{M.~Zechmeister\inst{1}\and A.~Reiners\inst{1}\and P.~J.~Amado\inst{2}\and
M.~Azzaro\inst{3}\and F.~F.~Bauer\inst{1}\and V.~J.~S.~B\'ejar\inst{4,5}\and
J.~A.~Caballero\inst{6}\and E.~W.~Guenther\inst{7}\and H.-J.~Hagen\inst{8}\and
S.~V.~Jeffers\inst{1}\and A.~Kaminski\inst{9}\and M.~K\"urster\inst{10}\and
R.~Launhardt\inst{10}\and D.~Montes\inst{11}\and J.~C.~Morales\inst{12}\and
A.~Quirrenbach\inst{9}\and S.~Reffert\inst{9}\and I.~Ribas\inst{12}\and
W.~Seifert\inst{9}\and L.~Tal-Or\inst{1}\and V.~Wolthoff\inst{9}}

\institute{Institut f\"ur Astrophysik, Georg-August-Universit\"at, Friedrich-Hund-Platz
1, 37077 G\"ottingen, Germany\\
\email{zechmeister@astro.physik.uni-goettingen.de}\and
Instituto de Astrof\'{\i}sica de Andaluc\'{\i}a (CSIC), Glorieta de la Astronom\'{\i}a s/n, 18008 Granada, Spain\and
Centro Astron\'omico Hispano-Alem\'an de Calar Alto (CSIC--MPG), Observatorio Astron\'omico Calar Alto s/n, Sierra de los Filabres-04550 G\'ergal (Almer\'{\i}a), Spain\and
Instituto de Astrof\'{\i}sica de Canarias, V\'{\i}a L\'actea s/n, 38205 La Laguna, Tenerife, Spain\and
Departamento de Astrof\'{\i}sica, Universidad de La Laguna, 38206 La Laguna, Tenerife, Spain\and
Departamento de Astrof\'{\i}sica, Centro de Astrobiolog\'{\i}a (CSIC--INTA), ESAC, Camino Bajo del Castillo, 28691 Villanueva de la Ca\~nada, Madrid, Spain\and
Th\"uringer Landessternwarte Tautenburg, Sternwarte 5, 07778 Tautenburg, Germany\and
Hamburger Sternwarte, Gojenbergsweg 112, 21029 Hamburg, Germany\and
Landessternwarte, Zentrum f\"ur Astronomie der Universit\"at Heidelberg, K\"onigstuhl 12, 69117 Heidelberg, Germany\and
Max-Planck-Institut f\"ur Astronomie, K\"onigstuhl 17, 69117 Heidelberg, Germany\and
Departamento de Astrof\'{\i}sica y Ciencias de la Atm\'osfera, Facultad de Ciencias F\'{\i}sicas, Universidad Complutense de Madrid, 28040 Madrid, Spain\and
Institut de Ci\`encies de l'Espai (IEEC-CSIC), Can Magrans s/n, Campus UAB, 08193 Bellaterra, Spain}

\date{Received / Accepted}

\maketitle

\section{Introduction}

The radial velocity (RV) method has been a very successful technique
to discover and characterise stellar companions and exo\-planets.
There are many algorithms to compute RVs, which can be
grouped depending on their complexity and choices for the model of
the reference spectra, number of model parameters, and statistics; these algorithms range
from simple cross-correlation with binary masks \citep{Queloz1995,Pepe2002},
to least-squares fit with coadded templates \citep{Anglada2012,Astudillo-Defru2015},
and finally to least-squares fitting with modelling of line spread
functions \citep{Butler1996,Endl2000}. Recently, Gaussian processes
have also been proposed to derive RVs \citep{Czekala2017}. The algorithm
choice is influenced by instrument type (e.g. those with stable point
spread functions) and the need for accuracy (e.g. fast and flexible
reduction directly at the telescope, generality) or high precision
\citep{David2014}. While most algorithms compute radial velocities
in the wavelength or pixel domain, there are also methods that work
in the Fourier domain (e.g. \citealp{Chelli2000}) and can efficiently
disentangle double-lined spectroscopic binaries (KOREL; \citealp{Hadrava2004}).
A review can be found in \citet{Hensberge2007}.

In this work we present our methods to measure RVs and spectral diagnostics,
which are implemented in a \textsc{Python} programme called SERVAL (SpEctrum
Radial Velocity AnaLyser). It aims for highest precision with stabilised
spectrographs. Therefore, we employ forward modelling in pixel space
to properly weight pixel errors, and we reconstruct the stellar templates
from the observations themselves to make optimal use of the RV information
inherent in the stellar spectra \citep{Zucker2003,Anglada2012}.

The SERVAL code was developed as the standard RV pipeline for CARMENES, which
consists of two high resolution spectrographs covering the visible
and near-infrared wavelength ranges from 0.52 to 1.71\,$\mu$m \citep{Quirrenbach2016}.
The CARMENES consortium regularly obtains spectra for about 300 M
dwarfs to search for exoplanets. The spectra are processed, wavelength
calibrated, and extracted by a pipeline called CARACAL \citep{Caballero2016}.
Using CARMENES and HARPS data we validate the performance of SERVAL,
which is publicly available%
\footnote{\url{www.github.com/mzechmeister/serval}}.

\section{\label{sec:RV-Method}Radial velocity algorithm}

Our RV computation aims for highest precision and is
based on least-squares fitting. \citet{Anglada2012} demonstrated
that this approach can yield more precise RVs than the cross-correlation
function (CCF) method. This is because the RV precision depends
on both the signal-to-noise ratio of the data and the match
of the model to the data. With a proper model of the observations,
the RV information, i.e. each local gradient, is optimally weighted
\citep{Bouchy2001} and outliers can be detected.

Our very general concept is to decompose simultaneously all observations
into (1)~a high signal-to-noise template $F(\lambda)$; (2)~RV shifts $v_{n}$ for observation number $n$; and (3)~multiplicative
background polynomials $p(\lambda)$ to account for flux variations,
e.g. wavelength-dependent fiber coupling losses due to imperfect correction
of the atmospheric dispersion. Therefore, our forward model $f(\lambda)$
for the observed flux is
\begin{equation}
f(\lambda)=p(\lambda)\cdot F(\lambda'(\lambda,v))\,,\label{eq:ymod}
\end{equation}
where the Doppler equation $\lambda'(\lambda,v)$ is given in Eq.~(\ref{eq:dopshift}).

We denote some $n=1\,...\, N$ observations taken at times $t_{n}$
each consisting of flux measurements $f_{n,i}$ at pixel $i$ with
calibrated wavelengths $\lambda_{n,i}$ and flux error estimates $\epsilon_{n,i}$.
For cross-dispersed echelle spectrographs the measurements are carried out
in several echelle orders $o$ (so a more explicit indexing is $f_{n,o,i}$).
The weighted sum of residuals is
\begin{align}
\chi^{2} & =\sum_{n,i}\frac{[f_{n,i}-f(\lambda_{n,i})]^{2}}{\epsilon_{n,i}^{2}}\label{eq:chi2_full}\\
 & =\sum_{n,i}w_{n,i}[f_{n,i}-p(\lambda_{n,i},a)\cdot F(\lambda'(\lambda_{n,i},v_{n}),b)]^{2}\label{eq:chi2_full_explicit}
\end{align}
with weights $w_{n,i}=\frac{1}{\epsilon_{n,i}^{2}}$, the polynomial
coefficients $a$, and the coefficients \textbf{$b$} describing the
template.

However, this simultaneous approach is hardly feasible in practice
because of the large amount of data (spectra, orders, pixels, $N\times O\times K$)
and large number of parameters. Therefore, we perform the decomposition sequentially
and iteratively. First, the observed spectrum with the highest signal-to-noise
is taken as a reference to shift all observations into this reference
frame and to coadd them to a high signal-to-noise template $F$ (Sect.~\ref{sub:Spectra-coadding}).
With this new template the radial velocities are recomputed while
the coefficients of the background polynomial are fitted simultaneously
(Sect.~\ref{sub:Least-square-RVs}). This one iteration is usually
sufficient \citep{Anglada2012,Astudillo-Defru2015}. Before we detail
the two steps in Sect. \ref{sub:Least-square-RVs} and \ref{sub:Spectra-coadding},
we define the model and prepare the data.

\subsection{Model details, data, and input preparation}

Given a template $F(\lambda')$ of the emitting source, a spectral
feature at reference wavelength $\lambda'$ in $F$ appears at $\lambda$
in the observation $f$, i.e. $f(\lambda)=F(\lambda')$.
The measured wavelength $\lambda$ depends on the
radial velocity of the moving source, i.e. the star, and on the velocity
of the moving observer on Earth \citep{Wright2014}.

To eliminate the contribution from Earth's motion, we transform the
measured wavelengths on Earth to the solar system barycentre via the
barycentric correction (cf. Eq.~(25) in with $\lambda_{\mathrm{meas}}=\lambda$ and $\lambda_{\mathrm{meas,true}}=\lambda_{B}$)
\begin{align}
\lambda_B & =\lambda(1+z_{B})\,,\label{eq:bary}
\end{align}
 where $z_{B}$ is the total redshift due to barycentric motion of
the observer. To compute the barycentric correction, SERVAL uses the
Fortran based code of \citet{Hrudkova2009}%
\footnote{\url{http://sirrah.troja.mff.cuni.cz/~mary/}%
}, which we modified to account for the proper motion of the star.
Moreover, we correct for secular acceleration \citep{Zechmeister2009},
which is in particular important for high proper motion stars and
requires the parallax of the star. For ultimate precision in the barycentric
correction, the code from \citet{Wright2014} with 1\,cm/s accuracy
can be used instead, which includes relativistic corrections%
\footnote{This requires internet access (\url{http://astroutils.astronomy.ohio-state.edu/exofast/barycorr.html})
or an installation of IDL.%
}.

\begin{figure*}
\includegraphics[width=1\linewidth]{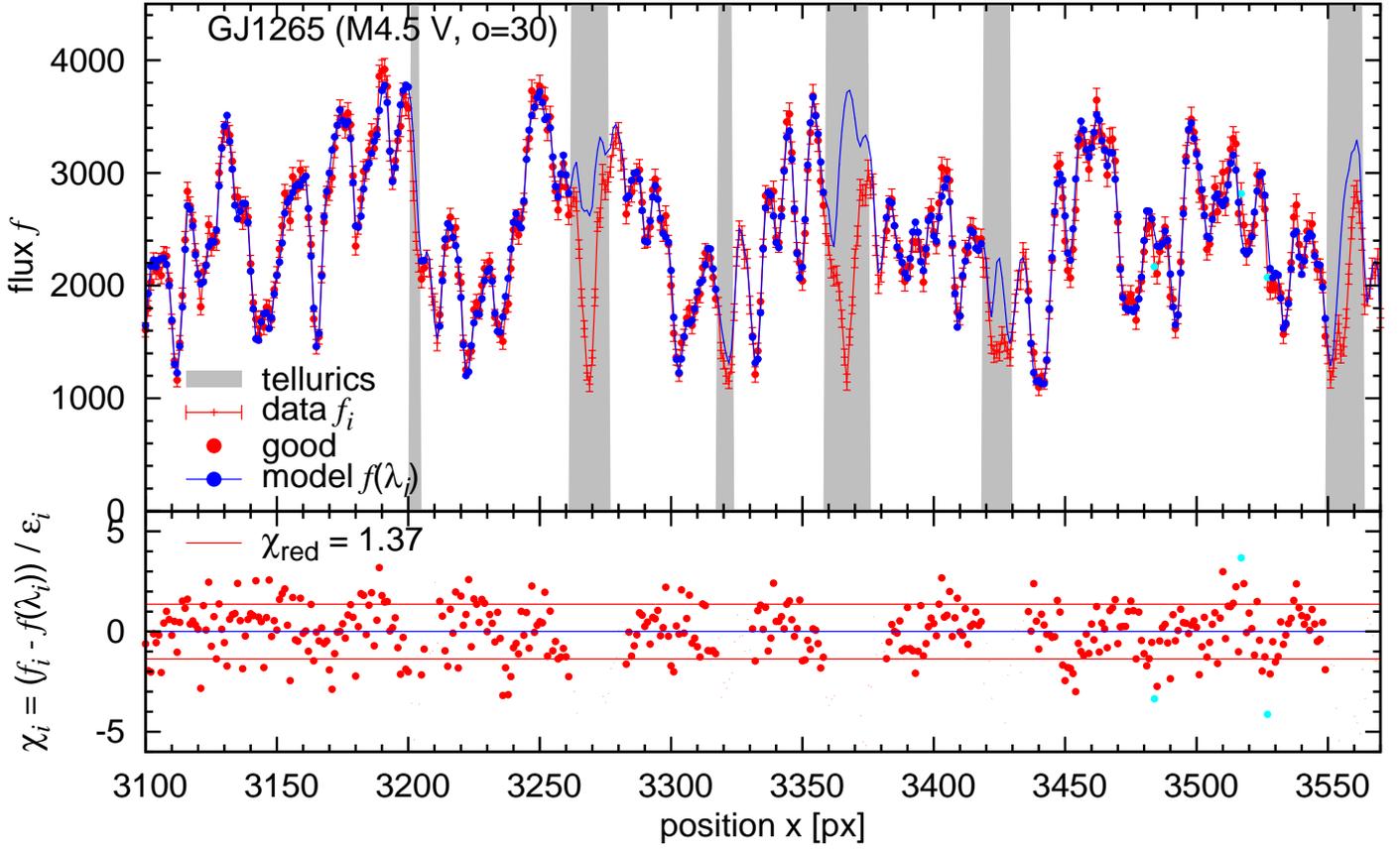}

\caption{\label{fig:Least-square-RVs}Least-squares RV fit for RV with CARMENES
VIS data for GJ~1265. Top: One observation (red) and the best fit
model (blue) are shown. Data points within telluric regions (grey area) are
a priori excluded. Data points classified as outliers are indicated in cyan.
Bottom: Fit residuals are shown.}
\end{figure*}

The stellar radial motion $v=cz$ redshifts the wavelength (cf. Eq.~(7)
in \citealp{Wright2014})
\begin{equation}
\lambda_B=\lambda'(1+z)\,,\label{eq:Doppler}
\end{equation}
where $c$ is the speed of light. Therefore, the equation
to Doppler shift the template with radial velocity parameter $v$
is
\begin{align}
\lambda'(\lambda,v) & = \lambda\frac{1+z_{B}}{1+\frac{v}{c}}\,.\label{eq:dopshift}
\end{align}
With the approximation $\frac{1}{1+\frac{v}{c}}\approx1-\frac{v}{c}$
and barycentric corrected wavelengths, one could write this as $\lambda'=\lambda_B\cdot(1-\frac{v}{c})$
as, for example, in \citet{Anglada2012}. The approximation is accurate to
1\,m/s over a velocity range of 17.3\,km/s, i.e. well sufficient
for the small amplitudes of exoplanets. But since there is actually
no need for this approximation, we employ the exact equation.

The SERVAL code has the option to perform the computation in logarithmic wavelengths.
Then Eq.~(\ref{eq:Doppler}) becomes
\begin{equation}
\ln\lambda_B=\ln\lambda'+\ln\left(1+\frac{v}{c}\right)\,.\label{eq:lnDoppler}
\end{equation}
For small shifts the Doppler shift is linear, $\ln\lambda_B\approx\ln\lambda'+\frac{v}{c}$.
This approximation is accurate to 1~m/s over a velocity range of
24.5\,km/s and should be avoided in high RV precision work in particular
in the barycentric correction in Eq.~(\ref{eq:bary}).

We assume that the spectrum is continuous and use cubic spline interpolation
to evaluate the template $F$ at any wavelength~$\lambda$, which
is needed for the forward modelling in Eq.~(\ref{eq:ymod}).

Furthermore, we generate and propagate a bad pixel map to flag pixels
with saturation, significant negative flux ($f_{i,n}<-3\epsilon_{i,n}$),
significant deviation in the fitting (outliers), or contamination
by tellurics and sky emission lines. The default telluric mask flags
atmospheric features deeper than 5\%. We also check the fits header
for the correct observing mode (e.g. accuracy and efficiency mode
in HARPS, dark time), available drift measurements, and suitable signal-to-noise
ratios. Those cases are flagged and, when required, excluded in the
analysis.

\subsection{\label{sub:Least-square-RVs}Least-squares RVs}

For the determination of the RV, we optimise Eq.~(\ref{eq:chi2_full_explicit})
with respect to the polynomial coefficients $a$ and the RV shift
$v$. The model is linear in $a$, but the RV parameter $v$ makes
the fit non-linear. There are several algorithms, e.g. Levenberg-Marquardt
or bisection, to solve non-linear least-squares problems. We use,
in analogy to the CCF method, a direct stepping through the velocity
parameter space $v$ with a default step size of $\Delta v=100$\,m/s
as a compromise between oversampling the resolution element (\textasciitilde{}km/s)
and computational speed. At fixed velocity $v_{k}$, the Doppler-shifted
template $F_{i,k}=F(\lambda_{i},v_{k},b)$ is evaluated at each pixel
$i$, and $\chi^{2}(v_{k})$ is obtained from a simple linear least-squares
fit for $a$
\begin{equation}
\chi_{k}^{2}=\sum_{i}w_{i}[f_{i}-p(\lambda_{i},a)\cdot F_{i,k}]^{2}\,.
\end{equation}
In the form $\chi^{2}=\sum_{i}F_{i}^{2}w_{i}[\frac{f_{i}}{F_{i}}-p(\lambda_{i},a)]^{2}$
and with the substitution $F_{i}^{2}w_{i}\rightarrow w_{i}$ and $\frac{f_{i}}{F_{i}}\rightarrow f_{i}$,
standard library routines for polynomial fitting can be applied (division
by zero flux $F_{i}$ needs to be handled). We include the
polynomial in the model. \citet{Anglada2012} argued that this ``would
couple the flux normalisation coefficients to Doppler factor in a
nonlinear fashion'' in their least-squares algorithm with iterative
differential corrections and, therefore, they applied it to the data.
However, they do not correspondingly re-adjust the errors as implied
in our derivation here. Since in practice the variations in the polynomial
function should be only a few percent, their approximation is feasible.

\begin{figure*}
\includegraphics[width=1\linewidth]{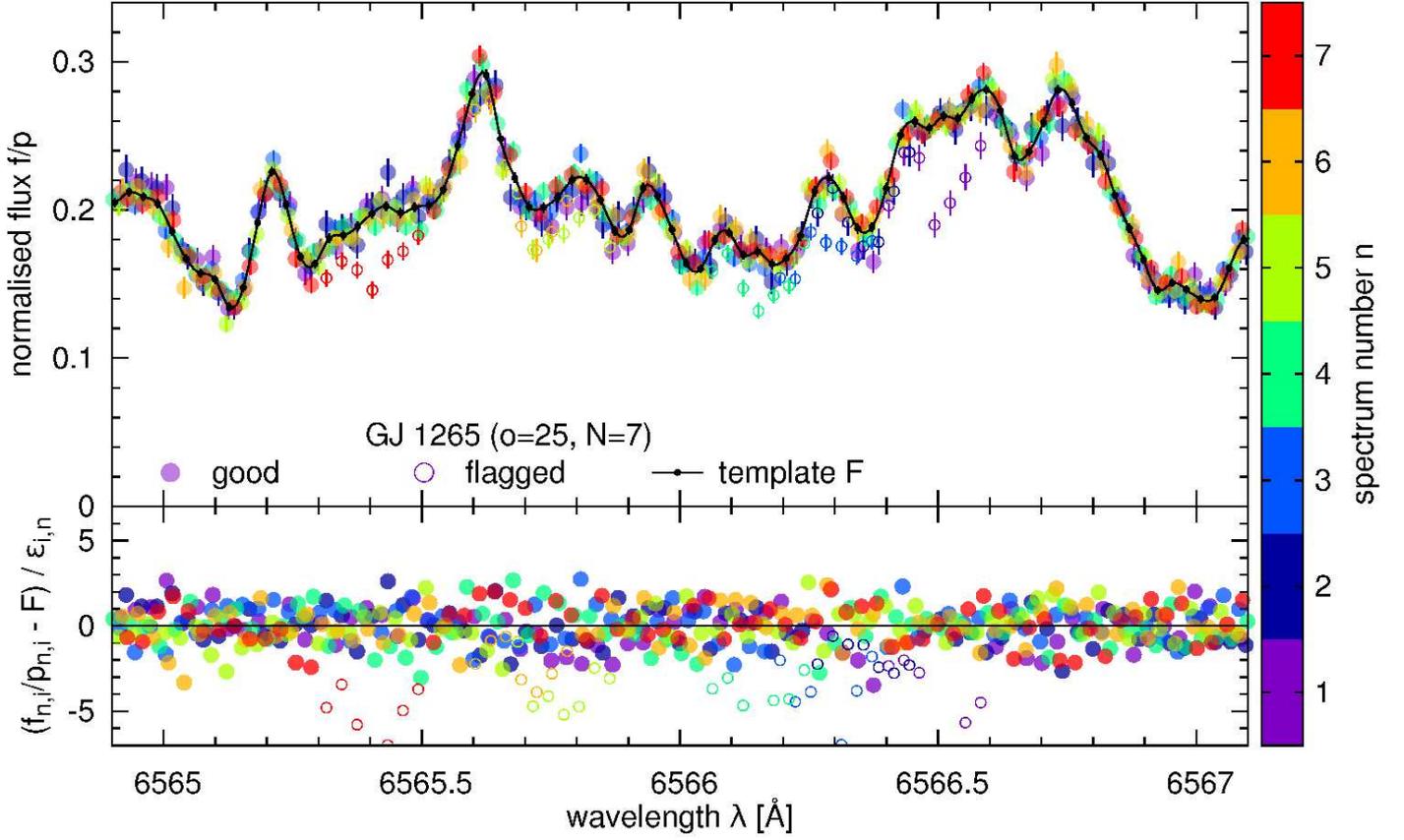}
\caption{\label{fig:Coadding}Spectra coadding of seven CARMENES VIS spectra
of GJ~1265. The seven observations (colour coded), each normalised
by a background polynomial, are fitted by a uniform cubic $B$-spline
(black) where basically each knot value is a free parameter (black
dots). Points affected by tellurics or outlying by 5$\sigma$ are
excluded from the fit (open circles).}
\end{figure*}

The $\chi^{2}(v)$ function sampled at $(v_{k},\chi_{k}^{2})$ is
explored for its global minimum. Around its minimum, the $\chi^{2}$
function takes a parabolic shape and the first derivative vanishes,
$(\chi^{2})'=0$. The Taylor expansion of the $\chi^{2}$ function
at the minimum to second order is
\begin{equation}
\chi^{2}(v+\delta v)\approx\chi_{\mathrm{min}}^{2}+(\chi^{2})'\delta v+\frac{1}{2}(\chi^{2})''\delta v^{2}=\chi_{\mathrm{min}}^{2}+\frac{1}{2}(\chi^{2})''\delta v^{2}\,.\label{eq:Taylor-expansion}
\end{equation}
Since we use cubic interpolation for the template, the second derivative
of the template $F$ (and of $\chi^{2}$) is continuous. Then a parabolic
interpolation through the minimum of the $\chi^{2}(v)$ function and
the two adjacent neighbours provides a refined estimate for $v$
and an error estimate for $v$ can be obtained from the parabola curvature.
The parabola minimum is located at the point,
\begin{equation}
v=v_{m}-\frac{\Delta v}{2}\cdot\frac{\chi_{m+1}^{2}-\chi_{m-1}^{2}}{\chi_{m-1}^{2}-2\chi_{m}^{2}+\chi_{m+1}^{2}}\,,\label{eq:RVo}
\end{equation}
where $m$ indexes the discrete global minimum of $\chi_{k}^{2}$ (see Eq.~(10.2.1) in
\citealp{Press1992} that is specialised here for a uniform grid, $\Delta v=v_{k}-v_{k-1}$,
or Eq.~\ref{eq:parabola_minimum} in Appendix~\ref{sec:Parabolic-interpolation}).

The uncertainty of $v$ is estimated from Eq.~(\ref{eq:Taylor-expansion})
with the $\Delta\chi^{2}(\delta v=\epsilon_{v})=\chi^{2}-\chi_{\mathrm{min}}^{2}=1$
criterion along with the curvature from Eq.~(\ref{eq:parabola_curvature})
\begin{equation}
\epsilon_{v}^{2}=2\frac{1}{(\chi^{2})''}=2\frac{\Delta v^{2}}{\chi_{m-1}^{2}-2\chi_{m}^{2}+\chi_{m+1}^{2}}\,.\label{eq:e_RVo}
\end{equation}
This value might be rescaled with $\chi_{\mathrm{red}}^{2}$ to account
for under- or overdispersion of the fit.

Finally, a statistical analysis of the fit is performed. The $\chi^{2}$
and $\chi_{\mathrm{red}}^{2}$ are computed. Linear residuals $r_{i}=f_{i}-f_{\mathrm{mod},i}$
exceeding the threshold
\begin{equation}
\left|r_{i}\right|>\kappa\epsilon_{i}\sqrt{\chi_{\mathrm{red}}^{2}}
,\end{equation}
where the clipping value $\kappa$ is typically 3...5, are flagged
as outliers (e.g. by setting their weights to zero $w_{i}=0$) and
excluded in a repeated fitting. The same can be formulated with normalised
residuals $\chi_{i}=\frac{r_{i}}{\epsilon_{i}}$
\begin{equation}
|\chi_{i}|>\kappa\sqrt{\chi_{\mathrm{red}}^{2}}\,.\label{eq:kappa_clip}
\end{equation}
Figure~\ref{fig:Least-square-RVs} illustrates the best fit between
one observation and a template.

Each order is fitted separately and finally a weighted mean for the
radial velocities $v_{o}$ from Eq.~(\ref{eq:RVo}) with errors $\epsilon_{v_{o}}$
from Eq.~(\ref{eq:e_RVo}) over all orders $o$ is computed (see
also Fig.~\ref{fig:chromatic-index})
\begin{equation}
v=\frac{\sum\epsilon_{v_{o}}^{-2}v_{o}}{\sum\epsilon_{v_{o}}^{-2}}\label{eq:RVmean}
\end{equation}
with the error estimate
\begin{equation}
\epsilon_{v}=\sqrt{\frac{1}{\sum\epsilon_{v_{o}}^{-2}}\cdot\frac{1}{N_{o}-1}\sum\frac{(v_{o}-v)^{2}}{\epsilon_{v_{o}}^{2}}}=\frac{\mathrm{wrms}}{\sqrt{N_{o}-1}}\label{eq:e_RVmean}
,\end{equation}
where wrms is the weighted root mean square and $N_{o}$ is the number
of orders.

\subsection{\label{sub:Spectra-coadding}Spectra coadding}

At first glance, coadding spectra seems straightforward. However,
the stellar spectra are Doppler shifted by potential Keplerian orbits
and by the barycentric motion of Earth. Thus, spectra from different
epochs are sampled at different wavelength footpoints. Hence naive
coadding would require either some kind of resampling or interpolation
of the observations onto a common wavelength grid (e.g. \citealp{Anglada2012}), which leads to difficulties in applying the data uncertainties, or
calculating some form of bin means, which ignores the local gradient
over the bin width. We point out that bin means are zeroth-order $B$-splines.

We carry out the ``co-adding`` with a uniform cubic basic spline
($B$-spline) regression to the normalised data. We emphasise
here the term regression, which means least-squares fit and should
be distinguished from spline interpolation and smoothing splines.
There are several benefits of this approach: (1) $B$-spline regression
is a linear least-squares method, and, therefore, it is fast; (2)
there is no need to interpolate the data; (3) data point uncertainties
can be easily taken into account; (4) robust statistics for outlier
detections can be obtained with kappa-sigma clipping; and (5) a spline
function is a direct outcome and consistent with our input for the
forward model.

Given $v_{n}$ for the $N$ observations from Sect.~\ref{sub:Least-square-RVs},
we recompute the polynomials $p_{n,i}$ for each order (now with
the mean RV $v_{n}$) to normalise the data ($\frac{f_{n,i}}{p_{n,i}}$)
and calculate the Doppler-shifted wavelengths with Eq.~(\ref{eq:Doppler}).
We factor $p_{n,i}^{2}$ in Eq.~(\ref{eq:chi2_full_explicit}) and
write it in the form
\begin{equation}
\chi^{2}=\sum_{n,i}p_{n,i}^{2}w_{n,i}\left[\frac{f_{n,i}}{p_{n,i}}-F(\lambda'_{n,i},b)\right]^{2}\,.\label{eq:chi2_template}
\end{equation}
Now the task is to find all the coefficients $b_{k}$ of the spline
with $K$ knots, so that the residuals are minimal. The knots of the
template are positioned in a uniform grid in (logarithmic) wavelengths
$\lambda_{k}$ ($\ln\lambda_{k}$) and $f_{k}$ (or $B$-spline coefficient
$b_{k}$, respectively). By default, the number of knots $K$ is similar
to the number of data points per spectrum in the order, i.e. we
have about one knot per pixel. Depending on the needs, the sampling
could also be decreased to obtain a smoother template (e.g. noisy
observations or fast rotators) or increased to obtain subpixel sampling
(many observations). It can be inferred from Eq.~(\ref{eq:chi2_template})
that the normalised error estimates are $\frac{\epsilon_{n,i}}{p_{n,i}}$.

Equation~(\ref{eq:chi2_template}) is a linear least-squares problem
due to the linearity of the coefficients in the $B$-spline \citep{deBoor1978,Dierckx1993,Eilers1996}
used to describe the template
\begin{equation}
F(x,b)=\sum b_{k}B_{k}(x)\,,
\end{equation}
where $B$ are the basic functions. Figure~\ref{fig:Coadding} shows
the normalised and Doppler-shifted data and the best spline fit.

However, caution is needed when there are large gaps in the data,
i.e. regions where we have no information to say anything about the
template (i.e. no constraints for $b_{k}$). Possible solutions are
(1) splitting the spline fit at those points; (2) fitting penalised
splines; or (3) heavily down-weight, but not fully reject, the flagged
data points (outliers, tellurics), which caused the gaps. We chose
the last option (down-weighting) for tellurics. This means that we
include regions heavily contaminated by telluric lines in coadding,
while in RV measurements they are totally excluded.

For each knot we estimate an error for the knot value as well as the
number of contributing good points. We calculate the errors in the
knot values by first estimating the error in the $B$-spline coefficients
as
\begin{align}
\epsilon_{b_{k}} & =1/\sqrt{\sum_{i}w_{i}B_{k}(x_{i})}
\end{align}
and then through error propagation
\begin{equation}
\epsilon_{f_{k}}=\frac{1}{6}\epsilon_{b_{k-1}}^{2}+\frac{4}{6}\epsilon_{b_{k}}^{2}+\frac{1}{6}\epsilon_{b_{k+1}}^{2}\,.
\end{equation}
This simplified estimate does not use the covariance matrix,
since calculating the inverse matrix would be very time consuming.

To increase the robustness against outliers, a few ($\leq$3) kappa-sigma
clipping iterations ($\kappa=5$) are performed for the coadding (similar
as in Eq.~(\ref{eq:kappa_clip})).

Problems can arise with this coadding method when the pixel phase
coverage is small because of a small barycentric range owing to close observations
or high ecliptic latitude. Such small coverage may lead to ringing in the spline
function. Reverting to simple bin means (zeroth order $B$-splines)
or using penalised splines (p-splines, as a generalisation of $B$-splines;
\citealp{Eilers1996}) may help in such situations. Also sharp (undersampled)
features, such as cosmics and emission lines, may lead to ringing features.
The SERVAL code offers the option to use p-splines, but the choice for the
value of penalty/smoothing parameter in each particular case will require some user testing or cross-validation algorithms. We also point out the Gaussian process approach of \citet{Czekala2017},
where the variance/smoothness parameter is a hyperparameter, and that
$B$-splines and p-splines can be considered as a special case of
Gaussian processes \citep{Rasmussen2006}.

SERVAL also offers the possibility to use external templates, e.g.
from other observed stars of similar spectral type or synthetic spectra.
The latter may be useful to derive absolute RVs and line indices (Sect.~\ref{sub:Spectral-indices}).

\section{Spectral diagnostic and activity indicators}

\subsection{Chromatic RV index}

\begin{figure}
\begin{centering}
\includegraphics[width=1\linewidth]{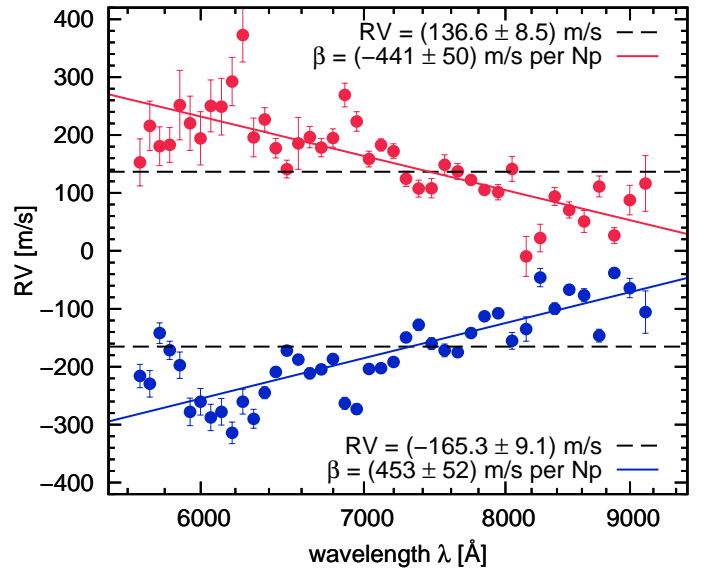}
\par\end{centering}

\caption{\label{fig:chromatic-index}Radial velocities measured in 42 orders of two CARMENES
VIS observations of YZ~CMi along with the simple weighted average
(dashed black line) and the chromatic index (solid blue and red line).
The wavelength axis is logarithmic.}
\end{figure}

In Eq.~(\ref{eq:RVmean}) we compute a simple weighted average for
the RV over the orders. However, since the echelle orders are related
to wavelength, we can also try to get some information about wavelength
dependency for instance by simply extending the model to a straight
line
\begin{equation}
v(o)=\alpha+\beta\ln\lambda_{o}\,,\label{eq:chromatic slope}
\end{equation}
where $\lambda_{o}$ is a representative wavelength of echelle order
$o$ (e.g. $\ln\lambda_{o}=\left\langle \ln\lambda_{o,i}\right\rangle $).
Via a $\chi^{2}$-fit we obtain a best fit estimate for the slope
parameter $\beta$, which we call chromatic index. Its unit is velocity
per wavelength ratio $e$ (Neper, symbol Np). Figure~\ref{fig:chromatic-index}
illustrates the slope definition for two observations of the active
M dwarf YZ~CMi.

\begin{figure*}[!t]
\includegraphics[width=0.33\linewidth]{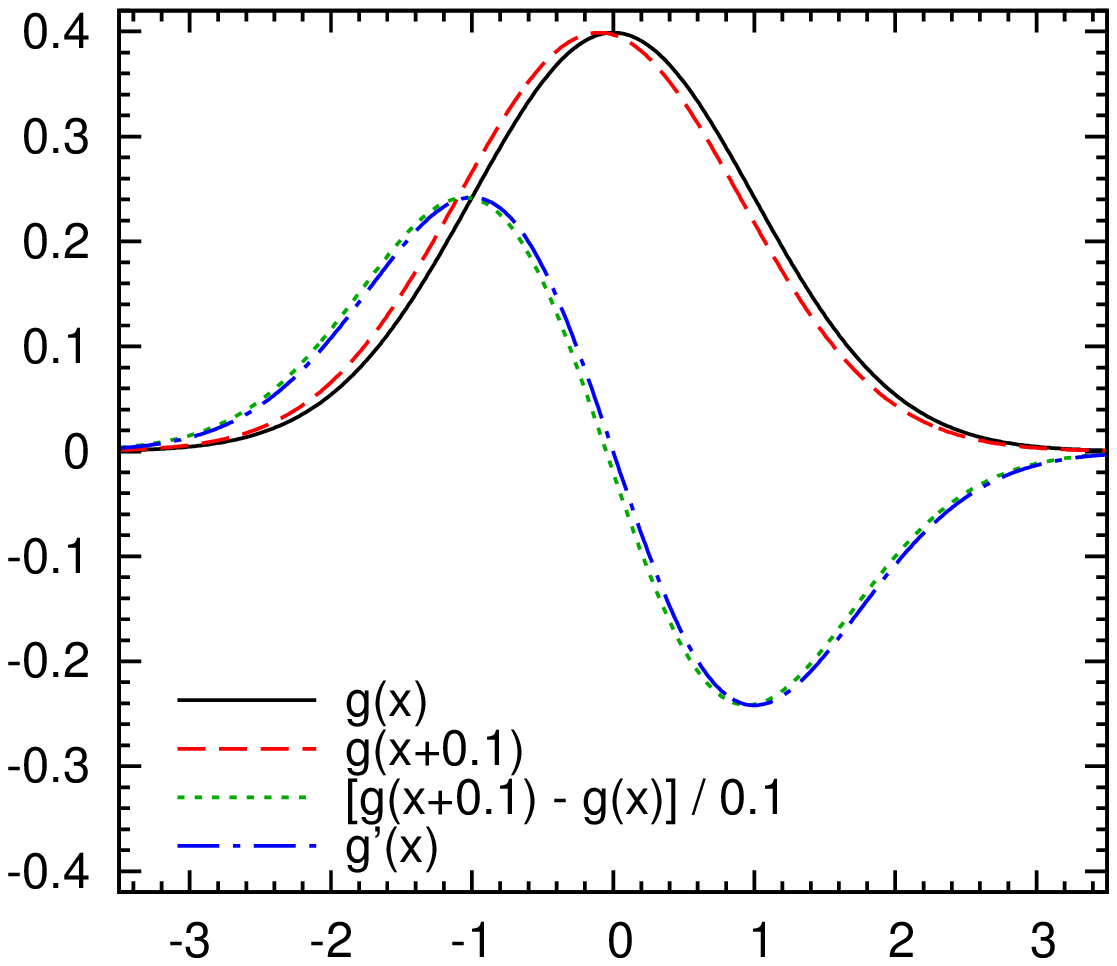}\includegraphics[width=0.33\linewidth]{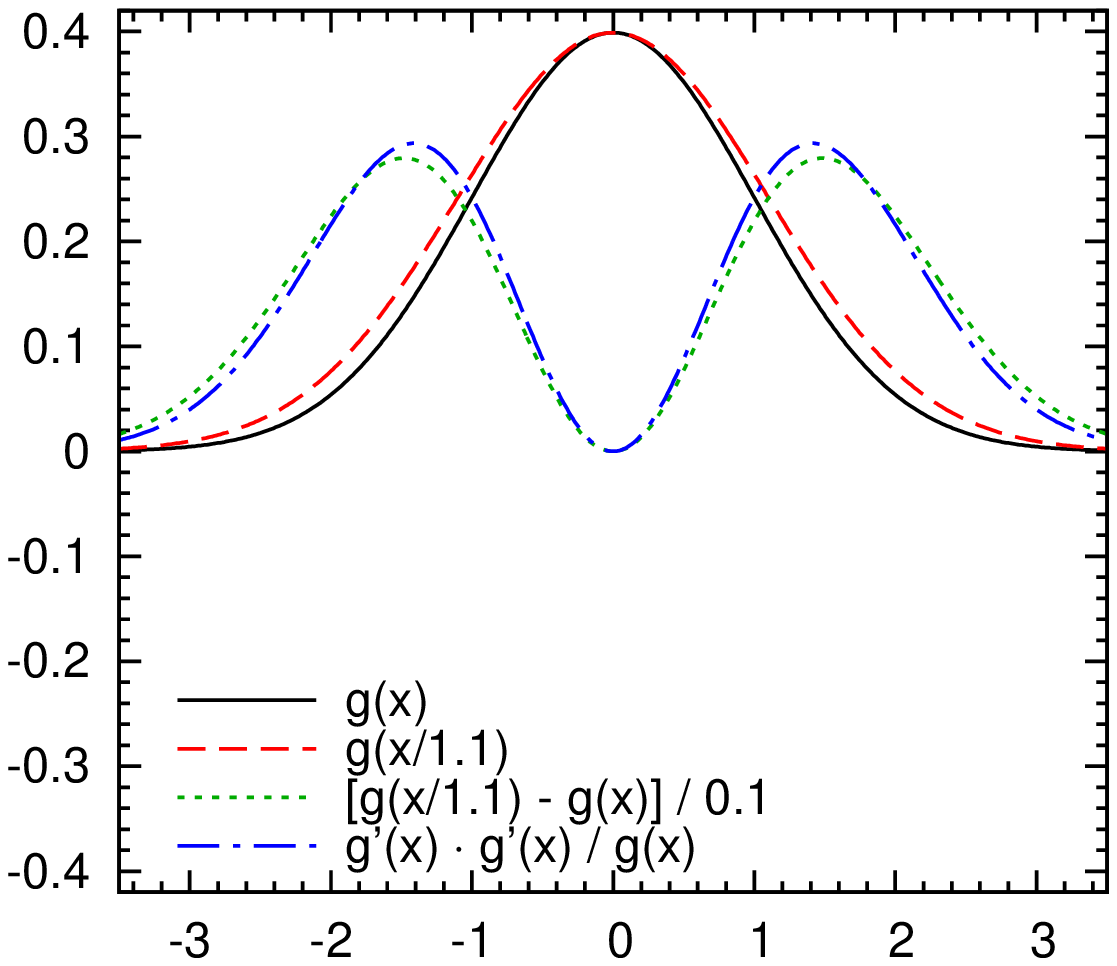}\includegraphics[width=0.33\linewidth]{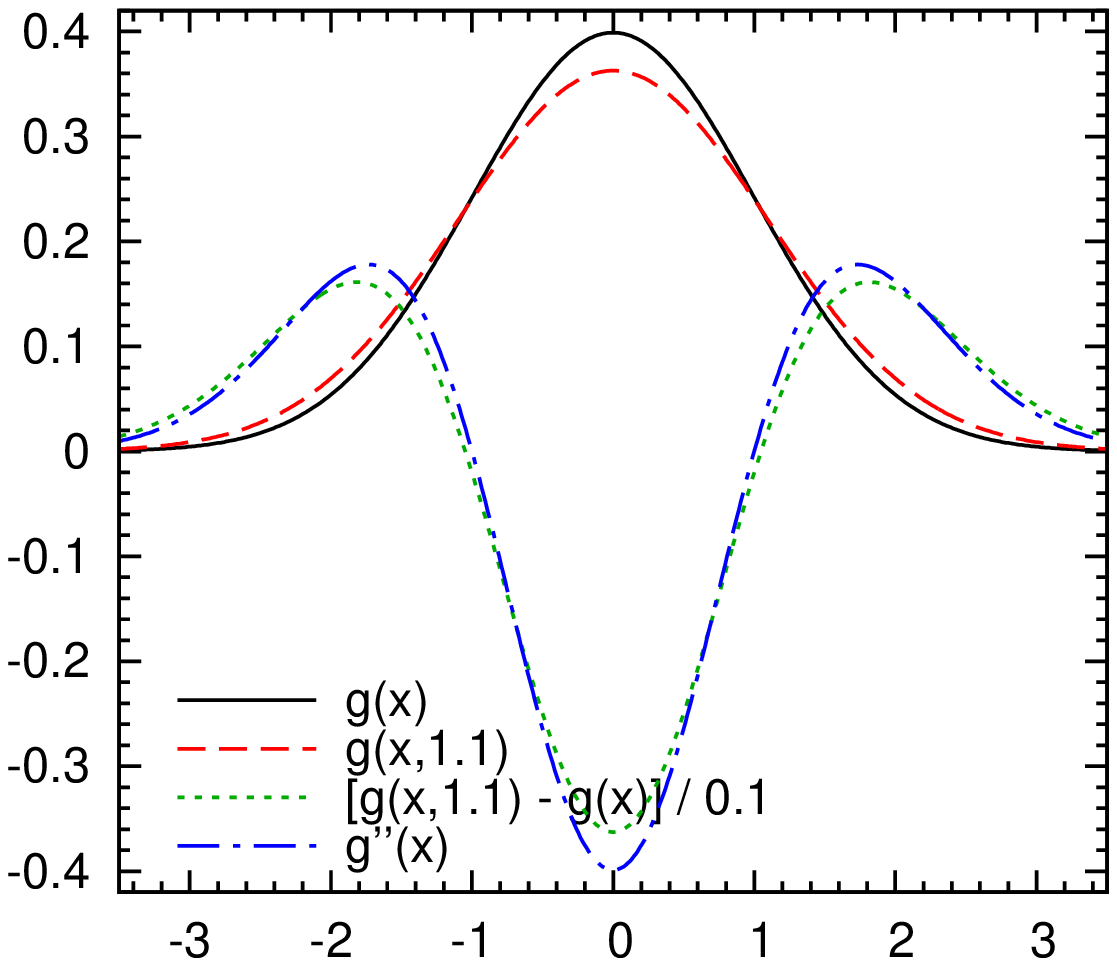}

\caption{\label{fig:dLW}Two Gaussian functions (solid black and dashed red)
with slightly different shifts (left panel) and slightly different
widths with same amplitude (middle panel) and same unit area (right
panel). Their scaled difference (dotted green) is related to the first
and/or second derivative (dash-dotted blue).}
\end{figure*}

The parameter $\alpha$ is considered as a nuisance parameter. Using
the mean velocity $v$ from Eq.~(\ref{eq:RVmean}) and the substitution
$\alpha=v-\beta\ln\lambda_{v}$, we can re-parametrise Eq.~(\ref{eq:chromatic slope})
as
\begin{equation}
v(o)=v+\beta\ln\frac{\lambda_{o}}{\lambda_{v}}
,\end{equation}
where $\lambda_{v}$ is the wavelength at which the slope intersects
the weighted mean RV. This allows us to report an effective wavelength,
which is in particular useful when comparing data from instruments
covering different wavelengths.

In the definition of the chromatic index in Eq.~(\ref{eq:chromatic slope}),
we chose $\ln\lambda$, the natural logarithm of wavelength $\lambda$,
as independent variable. While other choices like wavelength~$\lambda$
or order~$o$, which is basically equivalent to reciprocal wavelength~$\lambda^{-1}$
(i.e. frequency), lead to qualitatively very similar results for
$\beta$, our choice gives the same weights to each velocity element (resolution
element) and is also useful when comparing data from a variety of instruments.
Moreover, the wavelength range of 1 Np is typically covered by high
resolution echelle spectrographs, i.e. the values of the slopes are
of the order of the RV change over all echelle orders.

\subsection{\label{sub:dLW}Differential line width (dLW)}

A feature of the CCF method is that the CCF can be interpreted as
a mean stellar line profile that is actually convolved with a kind of kernel.
By analysing the CCF shape, for example by fitting a Gaussian function, we
can obtain information about the moments, among them the centre of
the Gaussian (first moment, i.e. RV) and the width (second moment,
i.e. full width half maximum, FWHM). To obtain information about asymmetries
(third moment, skewness), often bisectors are used (e.g. \citealp{Figueira2013}).

We study if there is an analogous set of parameters that can be associated
with least-squares fitting. Since the CCF is basically a $\chi^{2}$
function, one could consider fitting a Gaussian function to the $\chi^{2}$
function. However, we argue that the shape (curvature, asymmetry)
of the $\chi^{2}$ function is formed by the signal-to-noise of the
observation, masking of outliers and tellurics, and also the simultaneous
fit of the background polynomial.

In the following, we suggest another approach. We review
a simple method to measure differential RVs as implied in \citet{Bouchy2001}.
Figure~\ref{fig:dLW} (left) shows that when a Gaussian function
is slightly displaced, the residuals, i.e. the difference between
both curves, are correlated with the first derivative of the Gaussian
function. In fact, this is actually the definition of the first derivative.
From the scaling factor, which was already applied in Fig.~\ref{fig:dLW},
we can derive the RV
\[
g(x+\frac{v}{c})-g(x)\approx\frac{v}{c}\, g'(x)\,.
\]

Now we replace on the left side the function difference by the residuals
$r_{i}$ between (drifted) data and (flux-scaled) template, and on
the right side the derivative of the Gaussian function by the derivative
of the template. The derivative must be with respect to velocity,
i.e. $f'=\frac{\mathrm{d}f}{\mathrm{d}\ln\lambda}=\lambda\frac{\mathrm{d}f}{\mathrm{d}\lambda}$.
The relation becomes
\begin{equation}
r_{i}=\frac{v}{c}\, f'\!(x_{i})\,.
\end{equation}
In simple terms, we scale the first derivative to the residuals. The
best scaling factor $v$ can be derived with a linear least-squares
fit weighted with flux uncertainties $\sigma_{i}$. In Sect.~\ref{sub:Least-square-RVs},
on the other hand, the RV shift was already applied by actually minimizing
this correlation.

Now the question is whether we can find a similar relation using the
first or higher derivatives to check for variations in the line width.
Therefore, we vary the line width of the Gaussian. In one variant
we keep the line height constant (Fig.~\ref{fig:dLW}, middle) and
in another variant we keep the same area under the curve (Fig.~\ref{fig:dLW},
right). For both cases we plot again the scaled residuals and compare
these residuals with squared first derivative (middle) and second derivative
(right), respectively.

The case of unit area (right panel) results in balanced positive and
negative deviations, which is more similar to the situation we are
facing after a least-squares fit of the RV (Sect.~\ref{sub:Least-square-RVs}).
Indeed, we prove in Appendix~\ref{sec:Derivative-of-Gaussian} that
the mean value of the second derivative is zero (the same is true
for least-squares fit residuals) and that
\begin{equation}
g(x,\sigma+\delta\sigma)-g(x,\sigma)\approx\sigma\cdot\delta\sigma\cdot g''\!(x)\,.
\end{equation}
This relation is in a strict sense valid only for Gaussian functions,
but it suggests simply scaling the second derivative of the template
($f''=\frac{\mathrm{d}^{2}f}{\mathrm{d}\ln\lambda^{2}}=\lambda^{2}\frac{\mathrm{d}^{2}f}{\mathrm{d}\lambda^{2}}$)
to the residuals
\begin{equation}
r_{i}=\frac{\sigma\cdot\Delta\sigma}{c^{2}}\, f''\!(x_{i})\,.
\end{equation}
Hence we have to assume that the stellar lines are Gaussian-like shaped.
The relation also holds when we build up a spectrum by adding
Gaussian lines at other positions and with different strengths due
to the linearity. Then this also implies that it is applicable to
blended lines. However, the lines should have all the same width.
We point out that similar assumptions are made when fitting a Gaussian
profile to the CCF.

The scaling factor $\Delta\sigma$ carries information about line
width changes and we compute its value via a weighted linear least-squares
fit resulting in
\begin{equation}
\mathrm{dLW}\equiv\sigma\Delta\sigma=c^{2}\frac{\sum w_{i}f_{i}''r_{i}}{\sum w_{i}f_{i}''^{2}}
\end{equation}
and estimate its uncertainty through error propagation as
\begin{equation}
\epsilon_{\sigma\Delta\sigma}=c^{2}\sqrt{\frac{1}{\sum w_{i}f_{i}''^{2}}}\,.
\end{equation}
The dLW is computed in each order and finally averaged similar to
the RVs in Eq.~(\ref{eq:RVmean}) and (\ref{eq:e_RVmean}). The SERVAL
programme propagates the uncertainties of the spectra which are mostly photon and readout
noise. On top of this there can be other noise sources from the instrument,
such as focus or resolution change, or observation, such as line broading due
to barycentric motion during the exposure. Therefore, even quiet
stars with presumable stable intrisic line shapes may show excess
variations in line width indicators.

Since our template is a cubic $B$-spline, we can easily calculate
the second derivative $f''$ at each position of the spectrum.
Also the third derivative would be possible and the possibility of
defining a differential alternative to the bisector span in the CCF
would appear very straightforward, but this is not yet implemented.
Of course, the uncertainty in the higher order moments increase.

Owing to our differential approach, our width indicator $\sigma\Delta\sigma$
has also a differential nature, and, therefore, we call it differential
line width (dLW). Its unit is $\mathrm{m^{2}}/\mathrm{s}^{2}$ if
the derivative is calculated in velocity or logarithmic wavelength
scale. In analogy to differential RVs, in which an additive offset remains
with respect to to absolute RVs, the offset now becomes multiplicative
in the differential width (second moment). The $\sigma\Delta\sigma$
indicator  is sensitive to regions where the second derivative
is large, such as the cores of spectral lines (cf. Fig.~\ref{fig:dLW},
right panel).

Variations in the spectral line width can be intrinsic to the star, for example
pulsation or activity, or of instrumental origin in the form of focus change and smearing
due to the barycentric motion during exposure. In any case, correlation
of a RV signal with line width variations would argue against a planet
hypothesis.

\subsection{\label{sub:Spectral-indices}Line indices}

The SERVAL code also provides indices in the form of time series data for a
number of spectral lines (e.g. Ca~{\sc ii}~H\&K, H$\alpha$, Na~{\sc i}~D,
and Ca~{\sc ii}~IRT). Before measuring the line indices, we need to
find the line positions, i.e. the absolute RVs. These can be obtained
by measuring the RV of one spectrum against an absolute reference
(e.g. a PHOENIX spectrum).

Following \citet{Kuerster2003}, the indices are computed as
\begin{equation}
I=\frac{\left\langle f_{0}\right\rangle }{0.5\cdot(\left\langle f_{1}\right\rangle +\left\langle f_{2}\right\rangle )}\,,\label{eq:Line-index}
\end{equation}
where $\left\langle f_{0}\right\rangle $ is the mean flux around
the line centre and $\left\langle f_{1}\right\rangle $ and $\left\langle f_{2}\right\rangle $
are the mean fluxes over reference regions.

As an example, we choose for H$\alpha$ ($\lambda_{\mathrm{air}}=6562.8\,$\AA)
the region {[}-40,+40{]}\,km/s and as reference {[}-300,-100{]}\,km/s
and {[}+100,+300{]}\,km/s. We apply a larger core region compared
to \citet{Kuerster2003}, who only applied it to the inactive star
GJ~699, to collect all H$\alpha$ emission for most of the M dwarfs,
while for most active stars an even wider range could be considered
at the cost of decreasing index precision.

We estimate the error in the mean fluxes using the data error~$\epsilon_{i}$
\begin{equation}
\epsilon_{\left\langle f\right\rangle }=\frac{1}{N}\sqrt{\sum\epsilon_{i}^{2}}\,.\label{eq:e_mean_flux}
\end{equation}
This choice is not unique, but it preserves some information about
the signal-to-noise ratio in the spectrum; in case of pure photon
noise $\epsilon_{i}=\sqrt{f_{i}}$, it satisfies $\epsilon_{\left\langle f\right\rangle }=\sqrt{\frac{1}{N}\left\langle f\right\rangle }$.
From the stand point of least squares, the standard error of the mean
would be calculated via the standard deviation as $\frac{1}{\sqrt{N}}\sqrt{\frac{1}{N-1}\sum(f_{i}-\left\langle f\right\rangle )^{2}}$
and for a weighted mean as $\sqrt{\frac{1}{\sum\epsilon_{i}^{-2}}}$
(cf. Eq.~(\ref{eq:e_RVmean})). However, when modelling a line such
as H$\alpha$ with a mean, i.e. we fit a box, we should be aware of
a large model mismatch with this simplistic model making those error
estimates misleading. Indeed, one could consider more tailored models,
for example the Mount Wilson S-index employs a triangular shape \citep{Duncan1991}.

Finally, we simply propagate the errors from Eq.~(\ref{eq:e_mean_flux})
through Eq.~(\ref{eq:Line-index}) to an uncertainty for the line
index
\begin{equation}
\epsilon_{I}=I\sqrt{\frac{\epsilon_{0}^{2}}{\left\langle f_{0}\right\rangle ^{2}}+\frac{\epsilon_{1}^{2}+\epsilon_{2}^{2}}{\left\langle f_{1}\right\rangle ^{2}+\left\langle f_{2}\right\rangle ^{2}}}\,.\label{eq:e_Line-index}
\end{equation}

Sometimes (pseudo-) equivalent widths (pEW) are preferred instead
of line indices. We outlined in \citet{Zechmeister2009} that pEWs
are closely related to line indices, which might be interpreted as
``pseudo line heights''. Indeed, both quantities measure the zero
moment (area, integrated flux) and condense it into one parameter.
This simple estimate is sufficient for studies on the activity level
of stars and temporal variations. However, the true line shape is
more complex (e.g. self-absorption features) and needs to be described
by more moments/parameters.

We have also implemented a differential version for spectral line
indices where the mean fluxes in Eq.~(\ref{eq:Line-index}) are replaced
by the scaling factor between the observation and the coadded template.
This method aims for optimal weights and highest precision and would
be also applicable in case of missing data points, for example due to cosmics
or tellurics, but we recommend this only for inactive stars.

\section{Results}

\begin{figure*}
\begin{centering}
\includegraphics[width=0.33\linewidth]{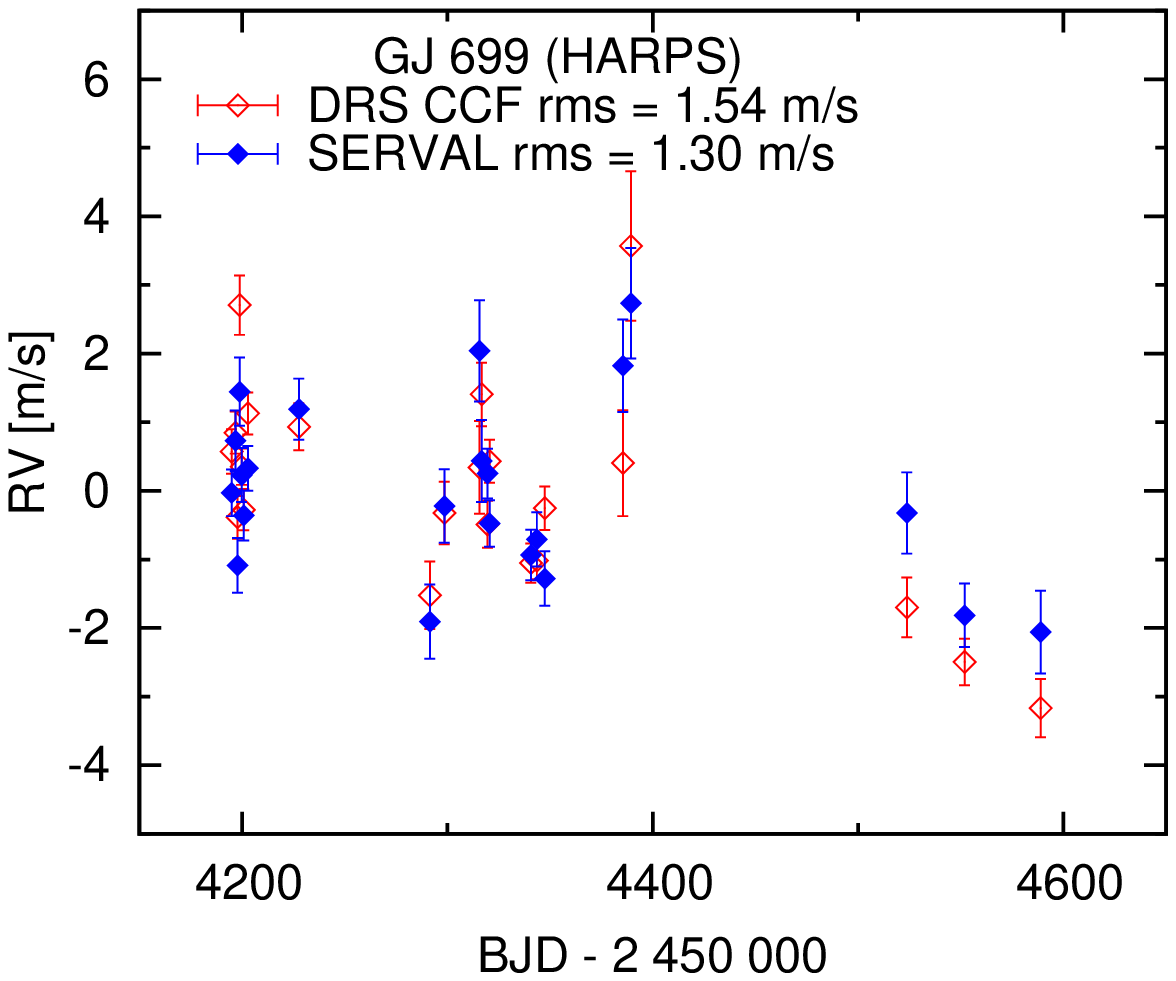}~\includegraphics[width=0.33\linewidth]{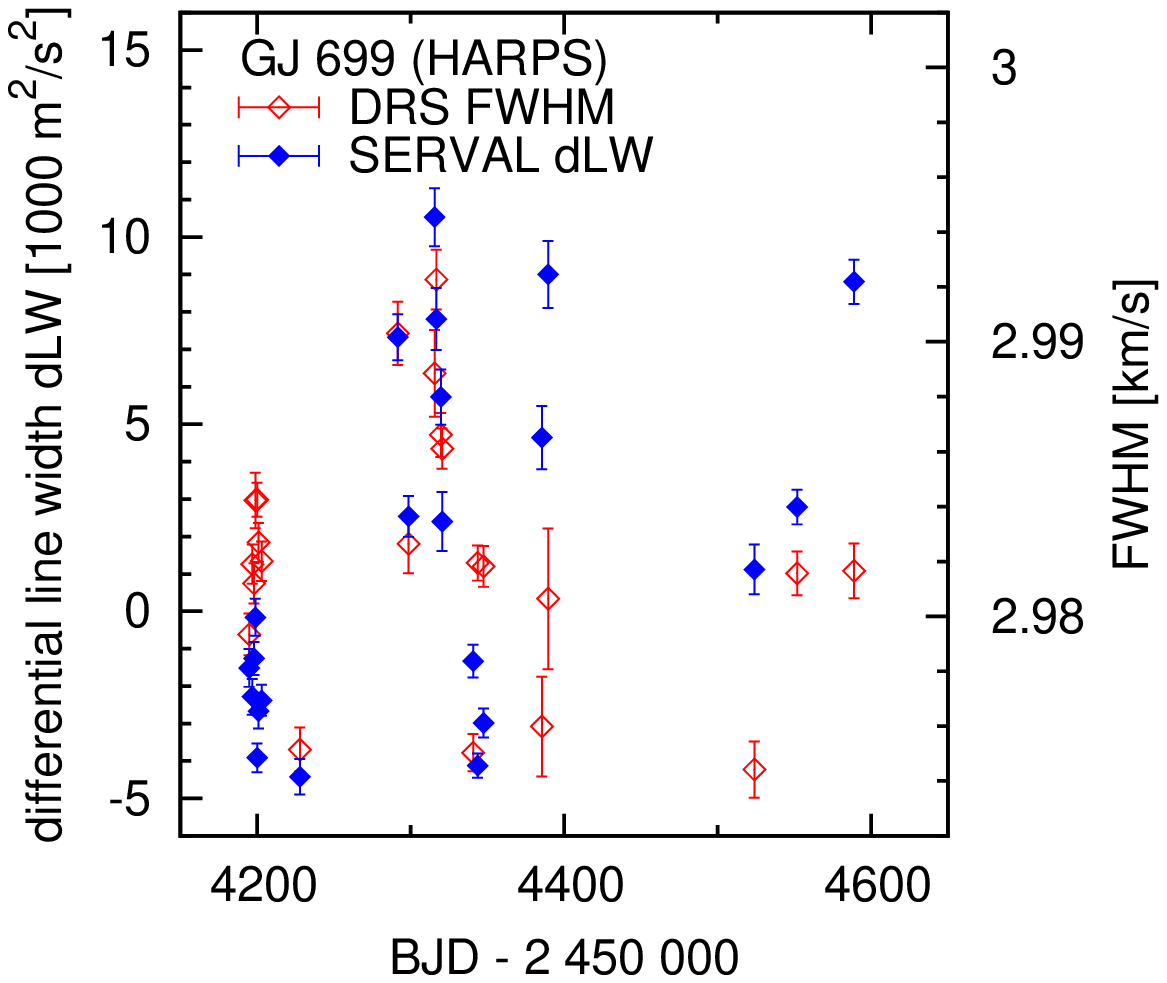}~\includegraphics[width=0.33\linewidth]{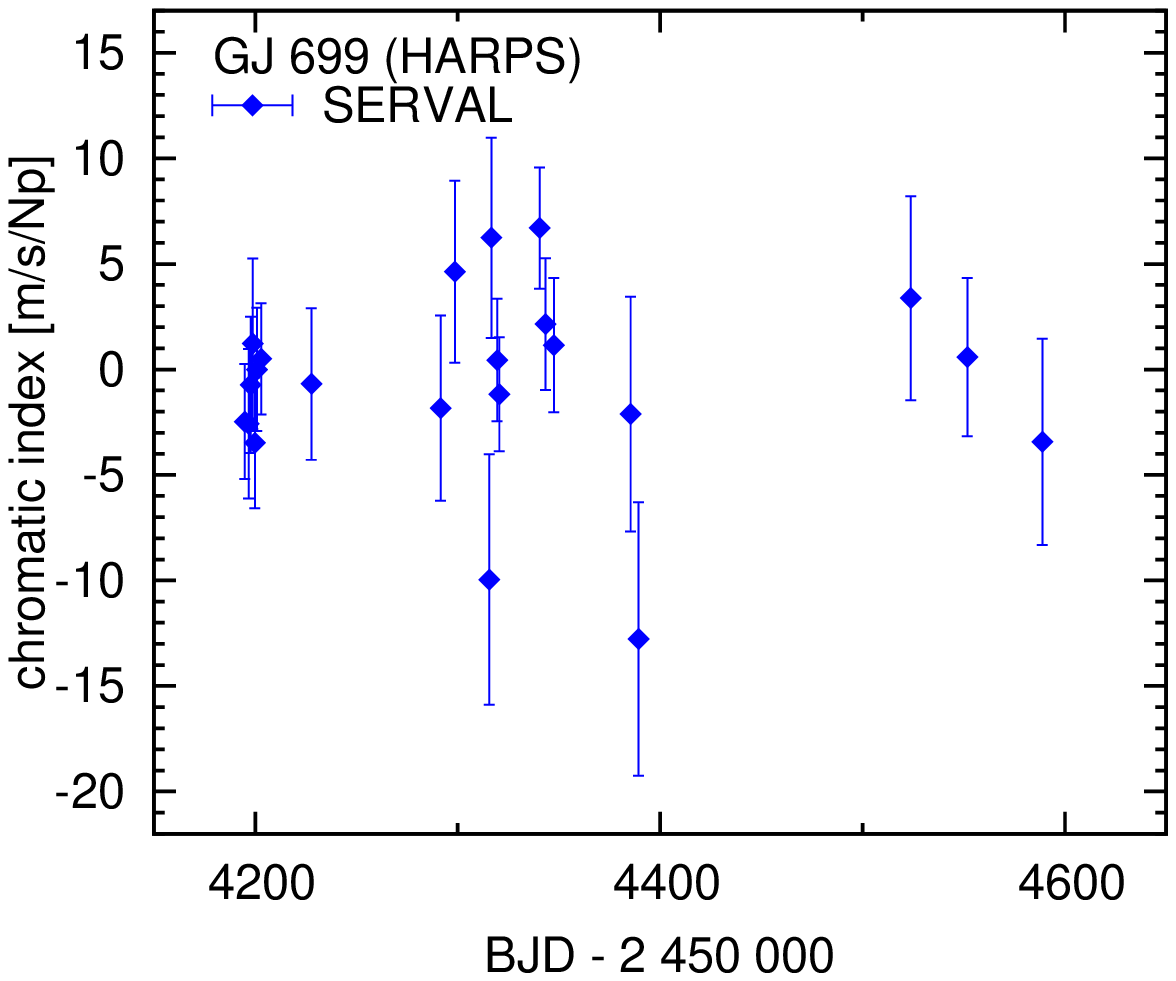}
\par\end{centering}

\caption{\label{fig:GJ699}Radial velocity (left), line width (middle), and chromatic index
(right) time series for HARPS data of GJ~699. SERVAL results are
indicated in blue solid diamonds and DRS in red open diamonds.}
\end{figure*}

\begin{figure*}
\begin{centering}
\includegraphics[width=0.33\linewidth]{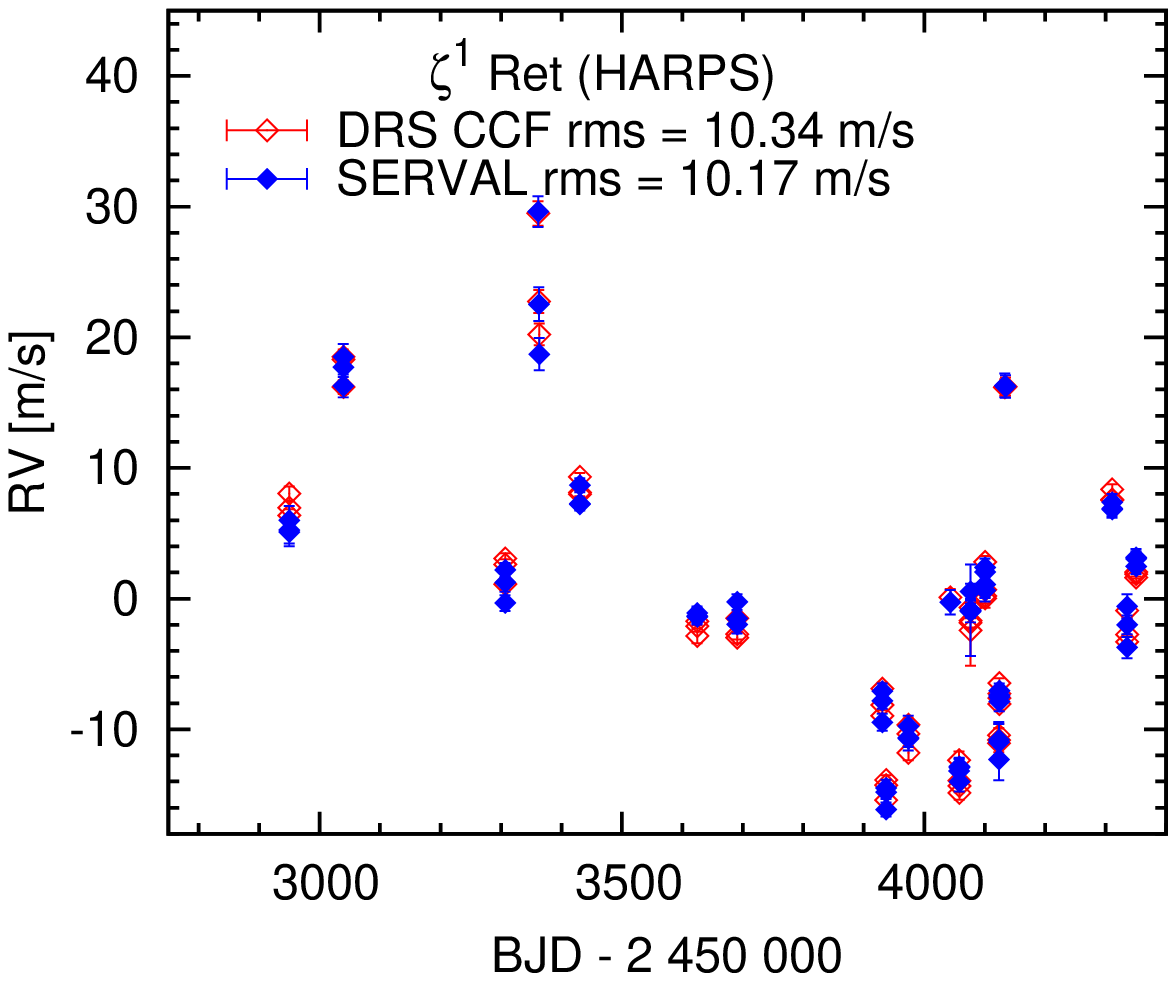}~\includegraphics[width=0.33\linewidth]{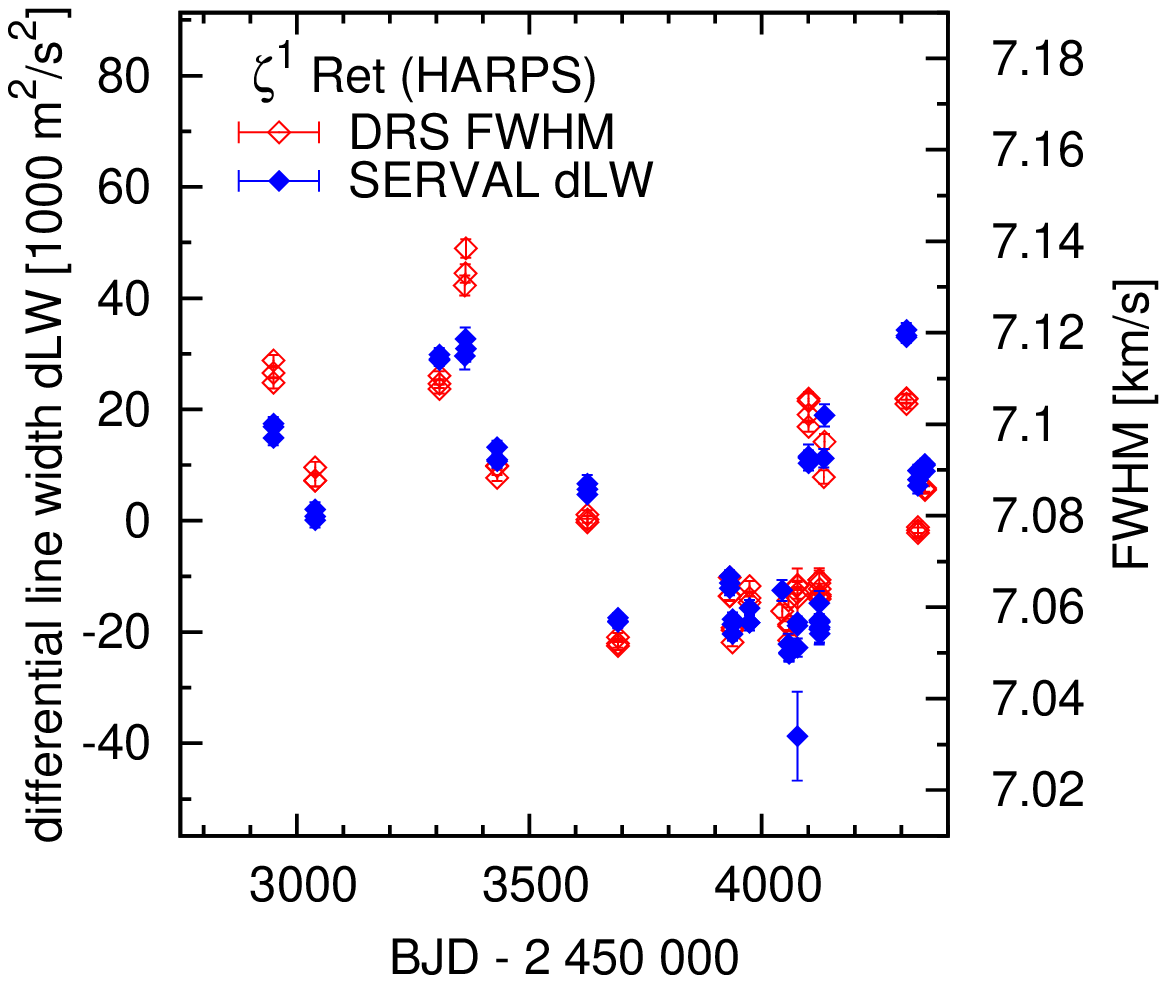}~\includegraphics[width=0.33\linewidth]{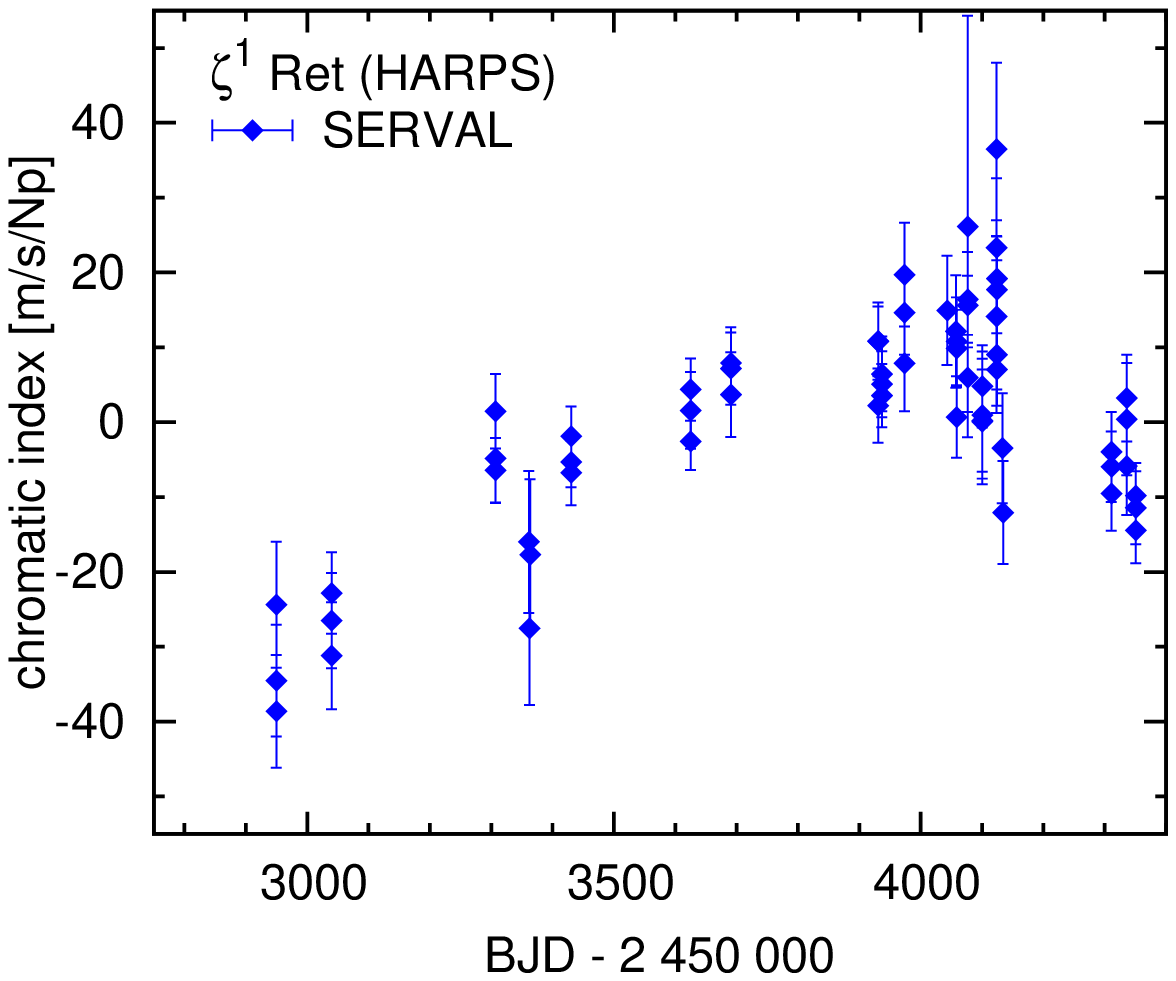}
\par\end{centering}

\caption{\label{fig:Zet1Ret}Same as Fig.~\ref{fig:GJ699} but for $\zeta^{1}$~Ret.}
\end{figure*}

\begin{figure*}
\begin{centering}
\includegraphics[width=0.33\linewidth]{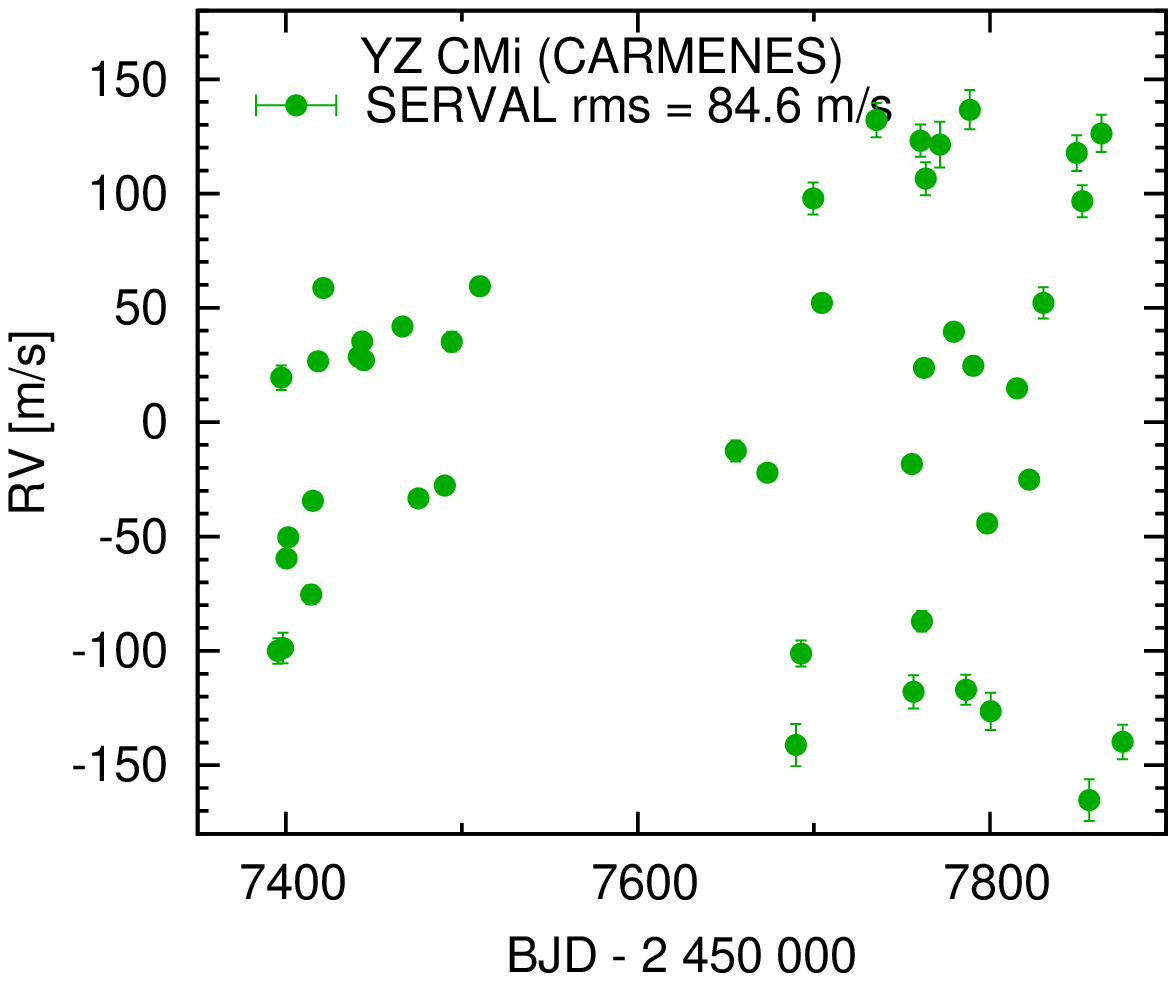}~\includegraphics[width=0.33\linewidth]{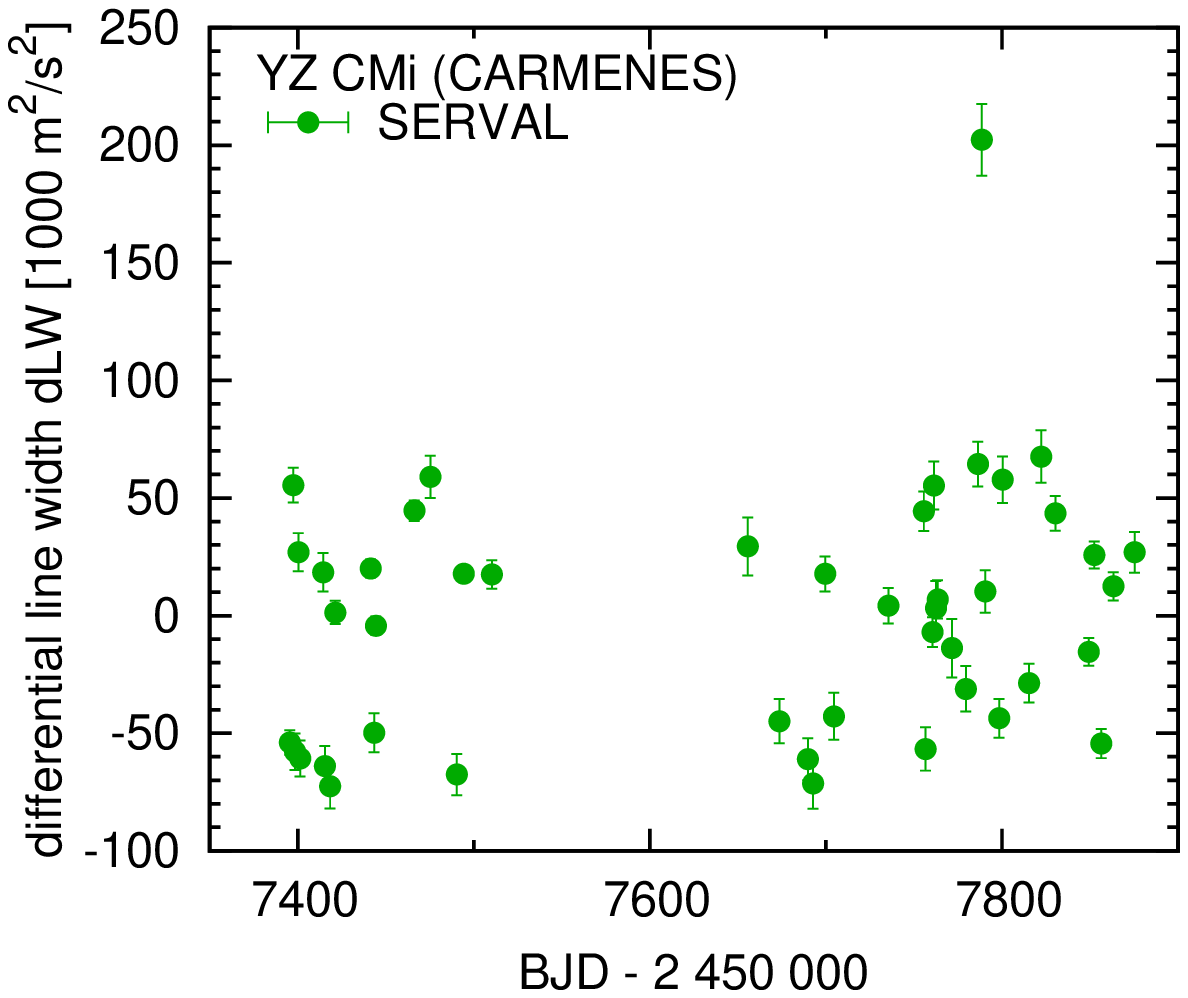}~\includegraphics[width=0.33\linewidth]{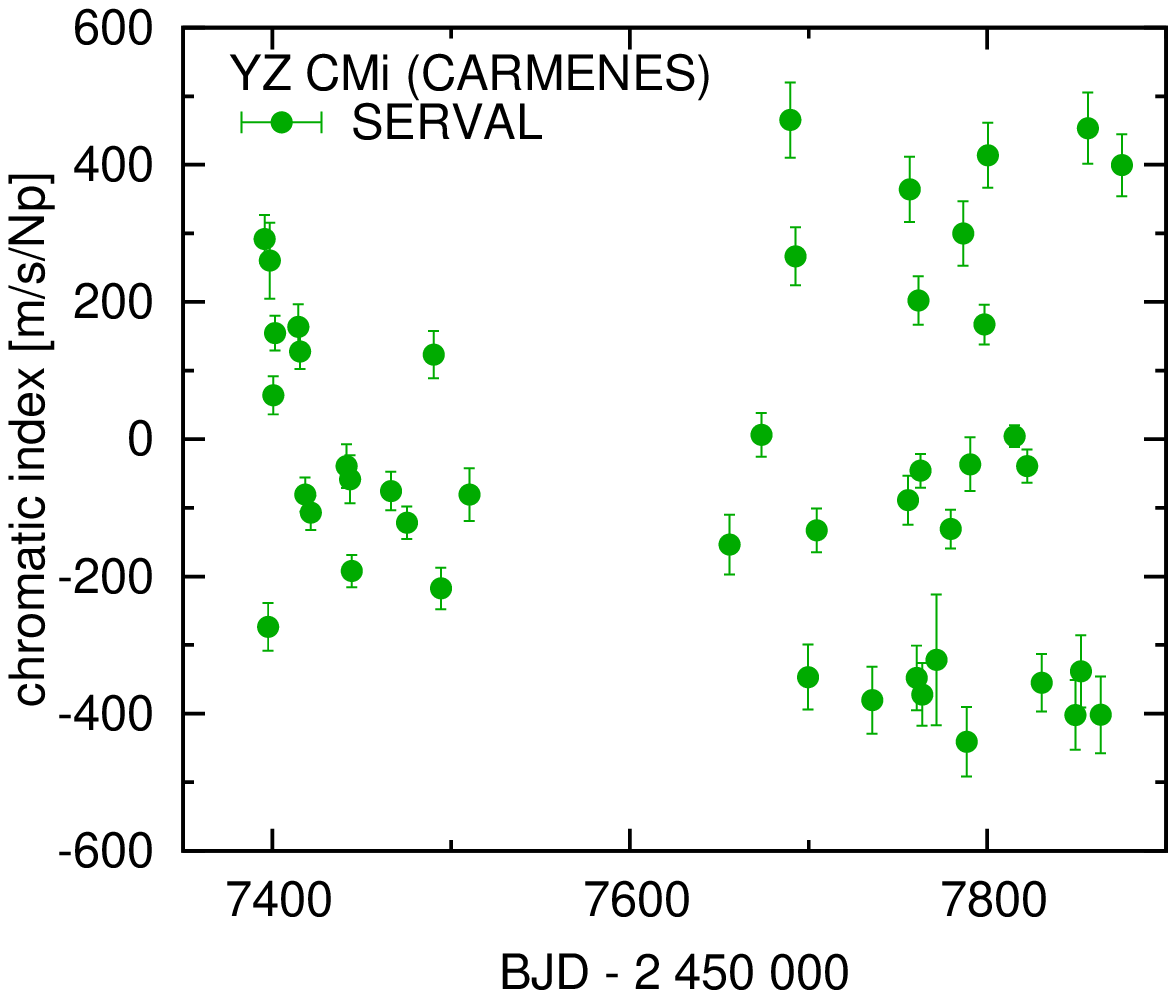}
\par\end{centering}

\caption{\label{fig:YZCMi}Radial velocity (left), dLW (middle), and chromatic index (right)
time series for CARMENES data of YZ CMi.}
\end{figure*}

\begin{figure*}
\begin{centering}
\includegraphics[width=0.33\linewidth]{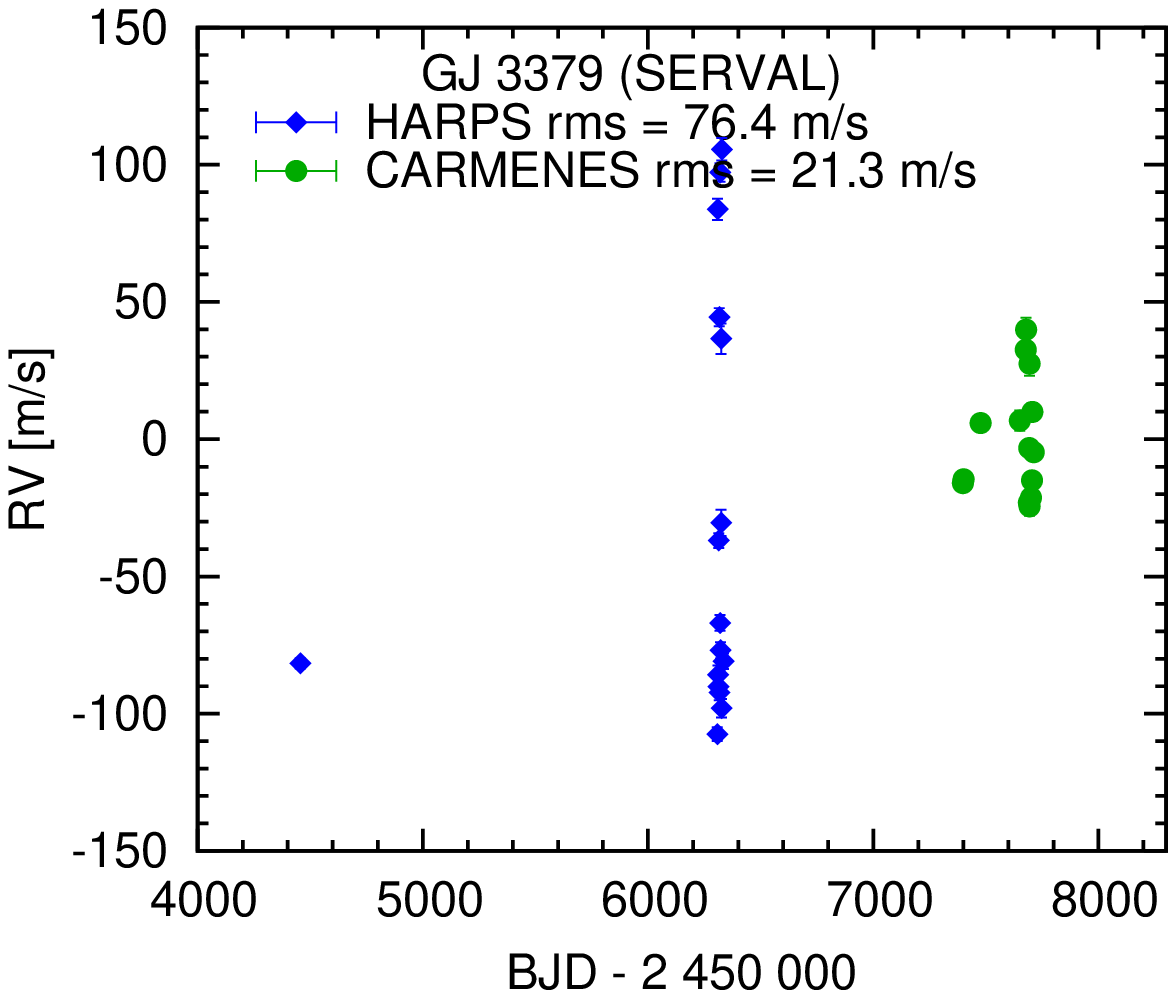}~\includegraphics[width=0.33\linewidth]{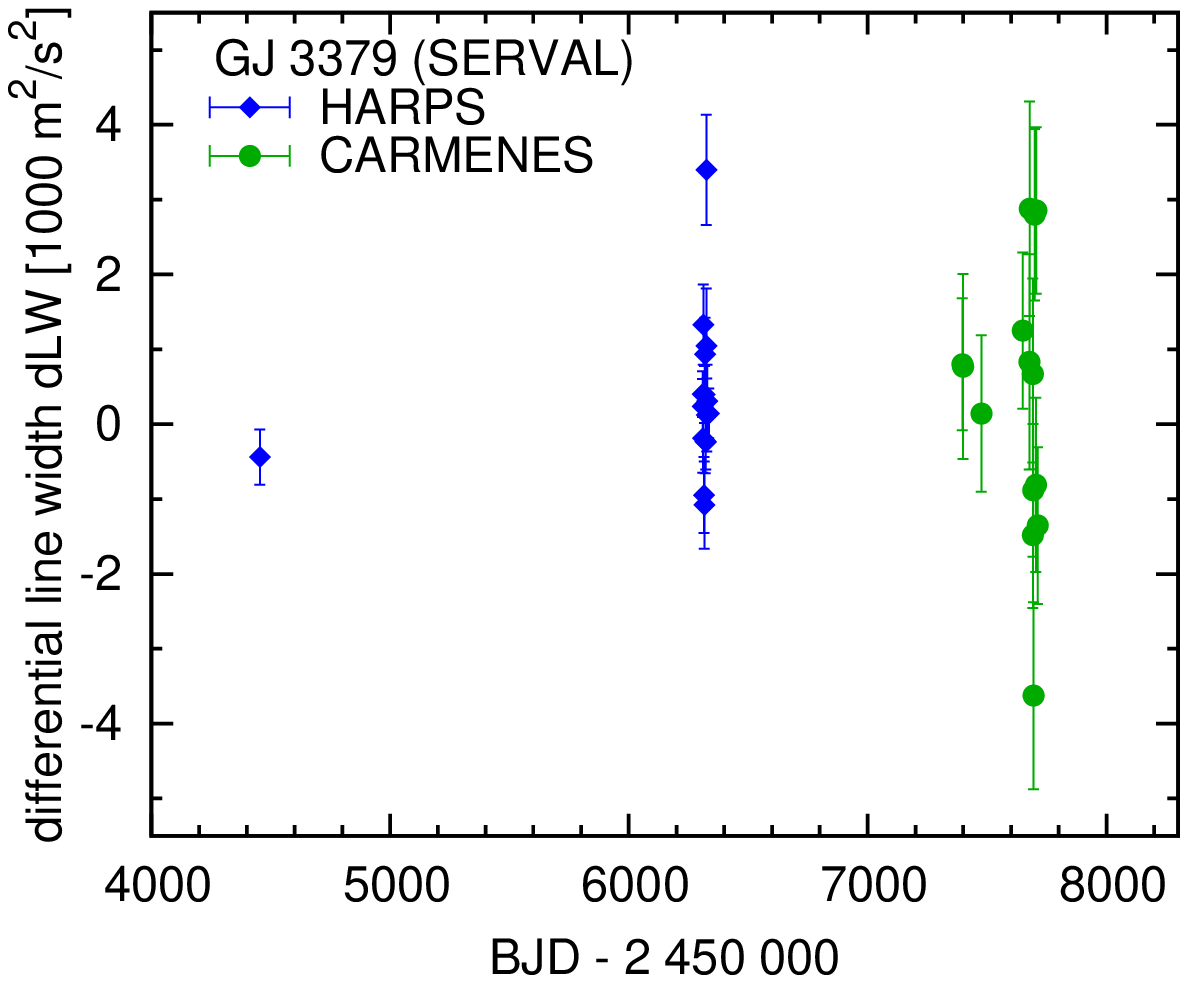}~\includegraphics[width=0.33\linewidth]{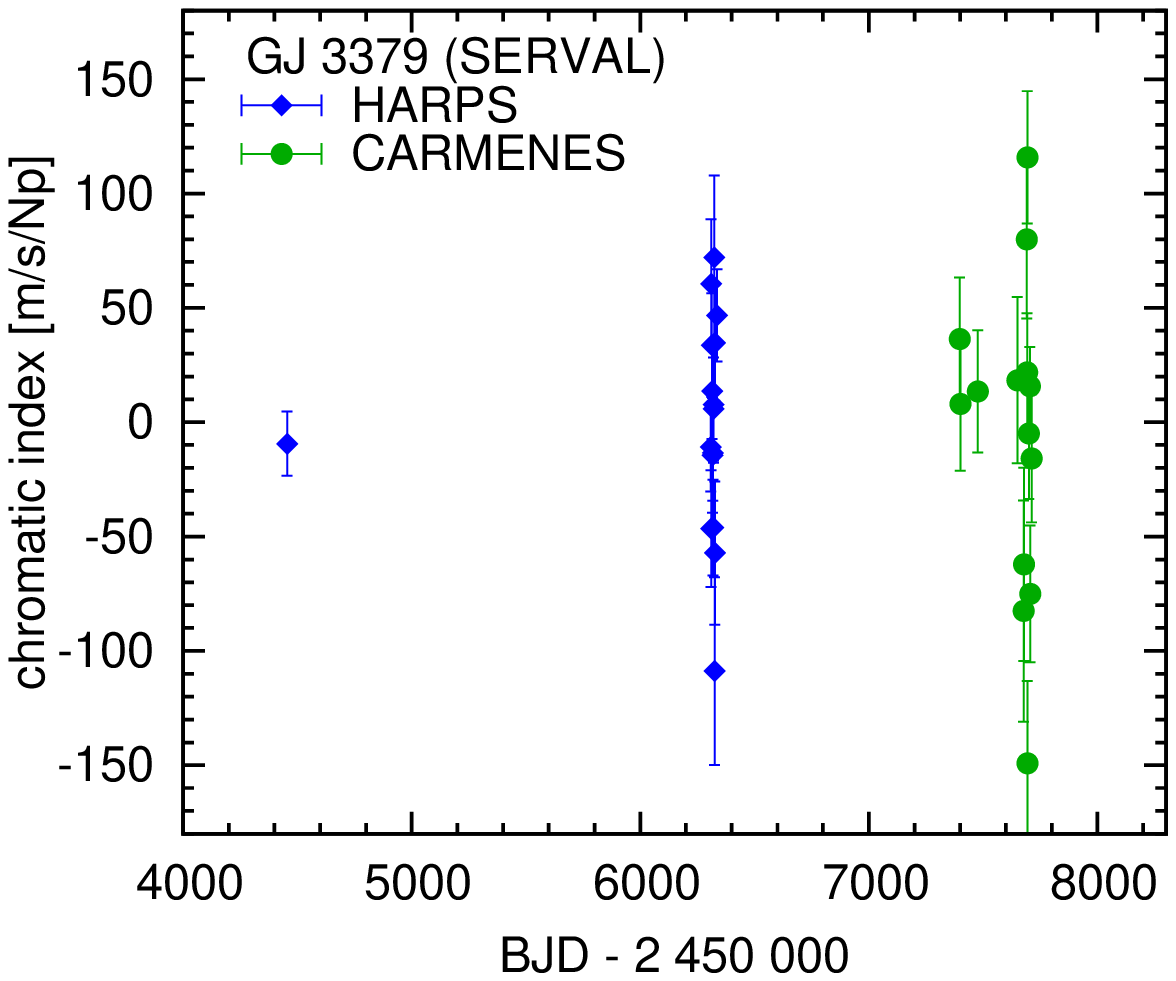}
\par\end{centering}

\caption{\label{fig:GJ3379}Same as Fig.~\ref{fig:YZCMi}, but for GJ~3379
and HARPS (blue diamonds) and CARMENES (green circles) data.}
\end{figure*}

In this section we evaluate the performance of SERVAL using its RV
precision, the dLW, and the chromatic index. To achieve this, we use
data from HARPS and CARMENES. The HARPS instrument and its data reduction
software has demonstrated long-term 1\,m/s precision since its installation
in 2003 \citep{Mayor2003}. Each of the points to be tested are evaluated
for (1) a stable and inactive M dwarf (Barnard's star), (2) a G4 dwarf
with strong correlations between FWHM and RVs ($\zeta^{1}$~Ret),
and (3) two very active M dwarfs (YZ CMi and GJ 3379). The time-series
measurements are shown in Figs.~\ref{fig:GJ699}--\ref{fig:GJ3379}.

\subsection{Targets}

\paragraph{GJ699} Barnard's star is an inactive M4V star. For comparison
we use the same 22 HARPS observations as previously used in \citet{Anglada2012}.
The spectra were obtained between 2007\nobreakdash-04\nobreakdash-04
and 2008\nobreakdash-05\nobreakdash-02. All spectra were secured
without simultaneous drift measurements and two spectra%
\footnote{2007-04-09T09:51:56.457 and 2007-04-10T09:55:45.767%
} were taken during twilight.

\paragraph{$\zeta^{1}$~Ret} This G4 dwarf was found by \citet{Zechmeister2013}
to have a strong correlation between RV and FWHM and hence we selected
it as an example for dLW analysis. We analysed 61 HARPS measurements
taken during 24 nights over a time span of 1401~days. Typical FWHM
values of 7.1\,km/s indicate that the star is not a slow rotator.

\paragraph{YZ~CMi} This star (GJ~285, Karmn~J07446+035) is a
very active M4.5 dwarf with significant rotation measured in several
works ($v\sin i\sim5$\,km/s, $\log(L_{H\alpha}/L_{{\rm bol}})=-3.48$;
\citealp{Reiners2012}; \citeauthor{Jeffers2017}), a photometric
period of 2.78\,d \citep{West2015}, an average magnetic field of
the order of 3--4\,kG, and local field strengths up to 7\,kG \citep{Shulyak2014}.
We obtained 45 CARMENES VIS observations spanning 480 days (from 2016\nobreakdash-01\nobreakdash-08
to 2017\nobreakdash-05\nobreakdash-01).

\paragraph{GJ~3379} This star (G~099-049, Karmn~J06000+027) is
an active M4.0 dwarf with significant rotational broadening ($v\sin i=(7.4\pm0.8)$\,km/s,
\citealp{Delfosse1998}). We found 16 HARPS spectra in the ESO archive
and obtained 14 spectra with CARMENES for GJ~3379. We only used this star to compare the chromatic index between HARPS and CARMENES wavelength
ranges.

\subsection{Criterion 1: Radial velocity precision}

\paragraph{GJ~699} Figure~\ref{fig:GJ699} (left) compares the
RVs derived with SERVAL (rms=1.30\,m/s) and the HARPS DRS pipeline
(rms=1.54\,m/s), while \citet{Anglada2012} reported a dispersion
of rms=1.24\,m/s for the same dataset. Secular acceleration is subtracted
in all cases. This demonstrates the capability of SERVAL to achieve
precise RVs at the 1\,m/s level.

\paragraph{$\zeta^{1}$~Ret} The RVs have an rms of 10 m/s (Fig.~\ref{fig:Zet1Ret}).
Since there is activity-induced jitter, we cannot evaluate the individual
RV performance with this star. Still, the RV difference between SERVAL
and DRS CCF has an rms of 1.2\,m/s and points to 1 m/s performance
of SERVAL for solar-like stars as well.

\begin{figure}
\begin{centering}
\includegraphics[width=0.8\linewidth]{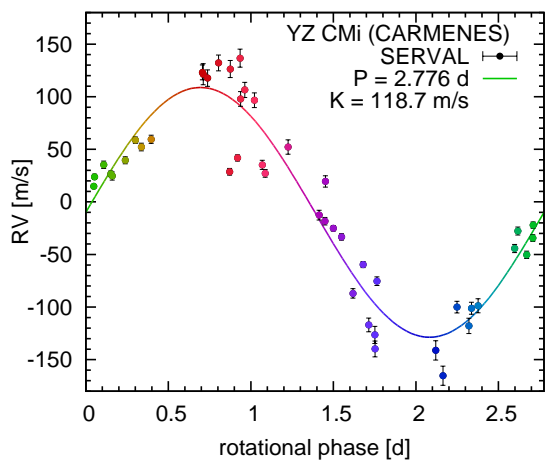}
\par\end{centering}

\caption{\label{fig:YZCMi-ph-RV}Radial velocities of YZ CMi phase folded to a period of
2.776\,d. The rotational phase is colour coded in reference to Fig.~\ref{fig:YZCMi-RV-csl};
blue and red correspond to the phases of predicted maximum
blue shift and red shift, respectively.}
\end{figure}

\paragraph{YZ~CMi} The RVs show an rms of 85\,m/s (Fig.~\ref{fig:YZCMi},
left). The variation takes place on short time scales (days) with
a significant perio\-di\-city of 2.776\,d with an amplitude of
120\,m/s (Fig.~\ref{fig:YZCMi-ph-RV} and a less significant one-day
alias at 1.556\,d and 100\,m/s). This period also appears in the
H$\alpha$-line equivalent width but at low significance.

\subsection{\label{sub:Zet1Ret}Criterion 2: dLW}

\begin{figure}
\begin{centering}
\includegraphics[width=0.5\linewidth]{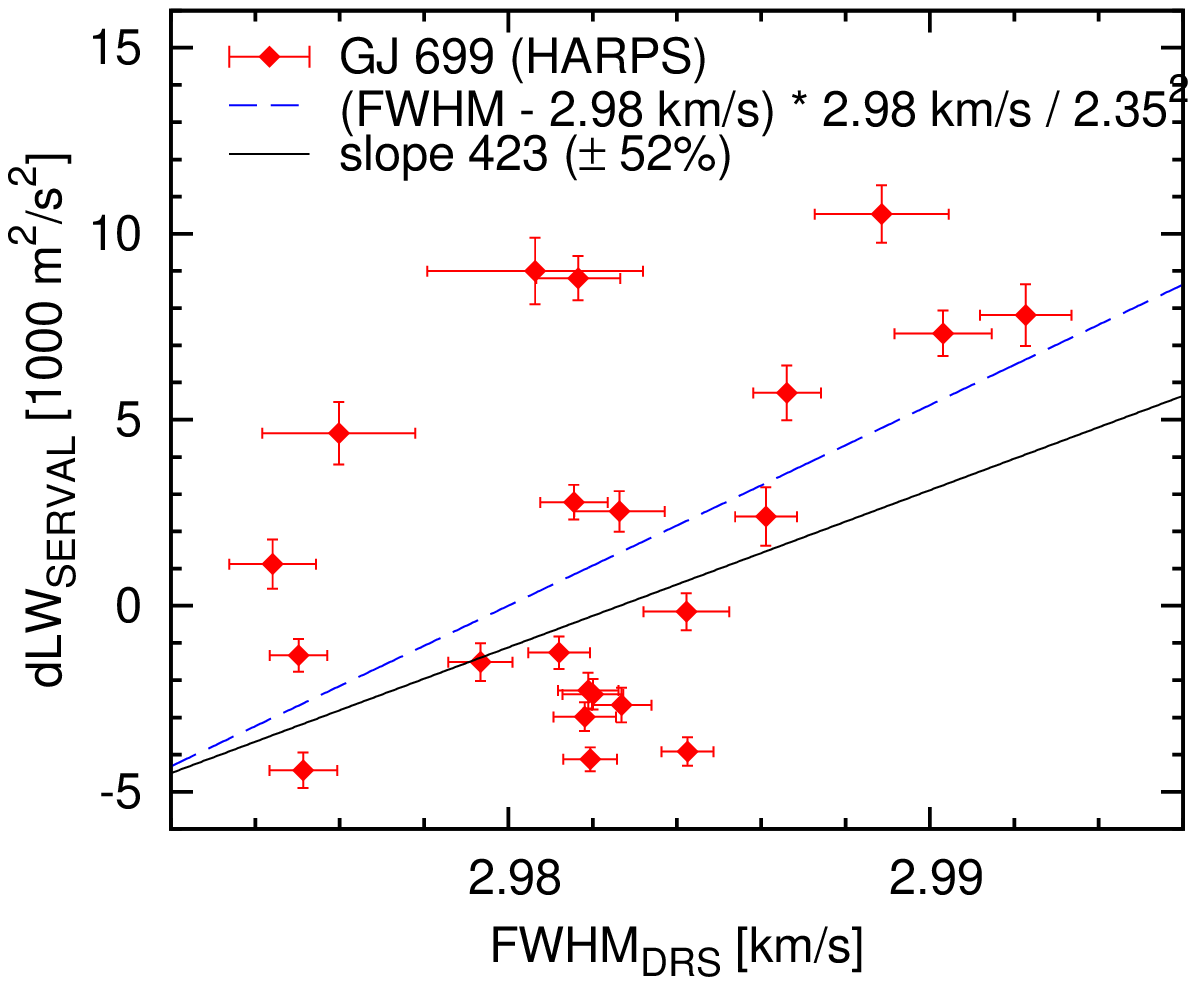}\includegraphics[width=0.5\linewidth]{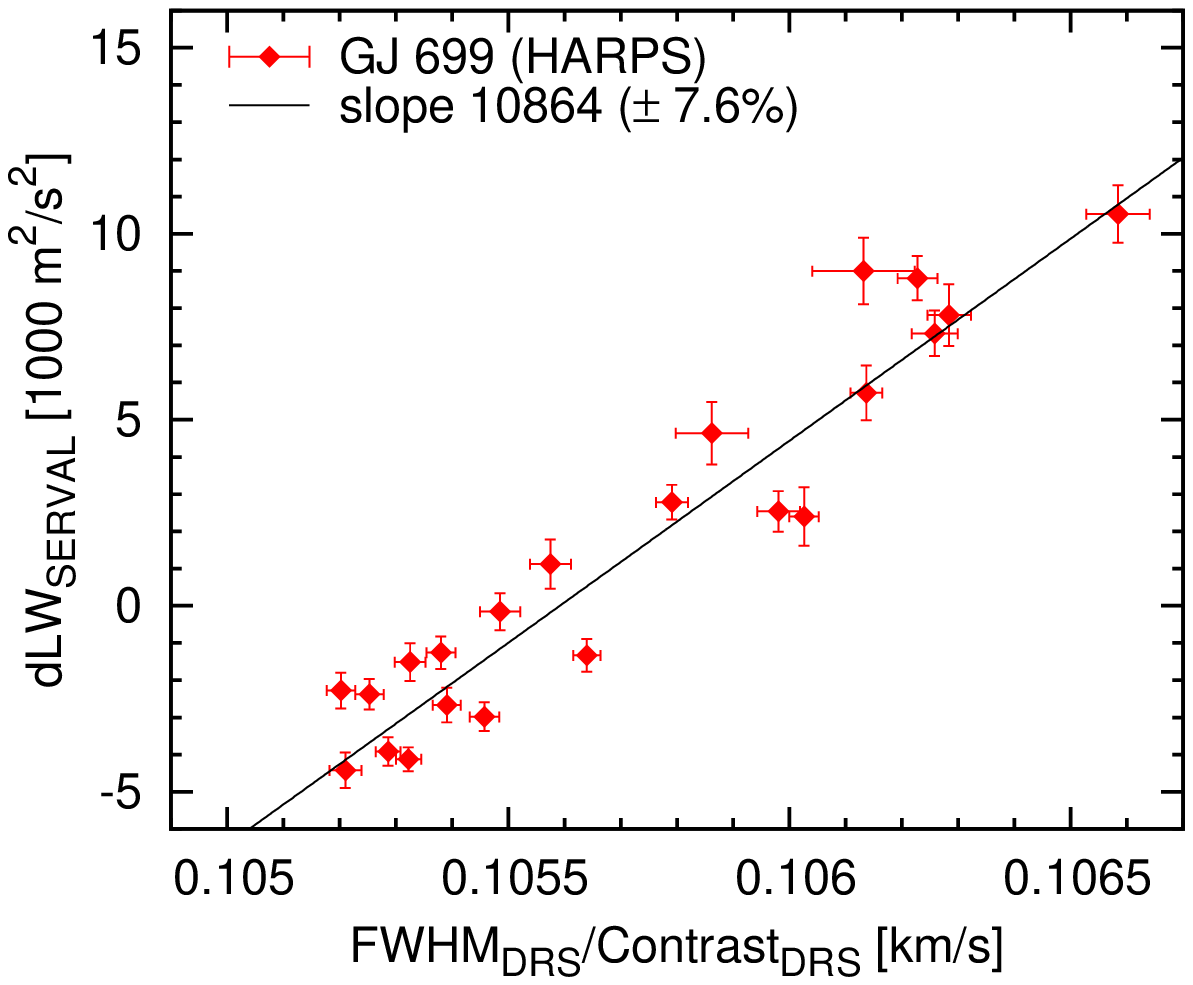}
\par\end{centering}

\begin{centering}
\includegraphics[width=0.5\linewidth]{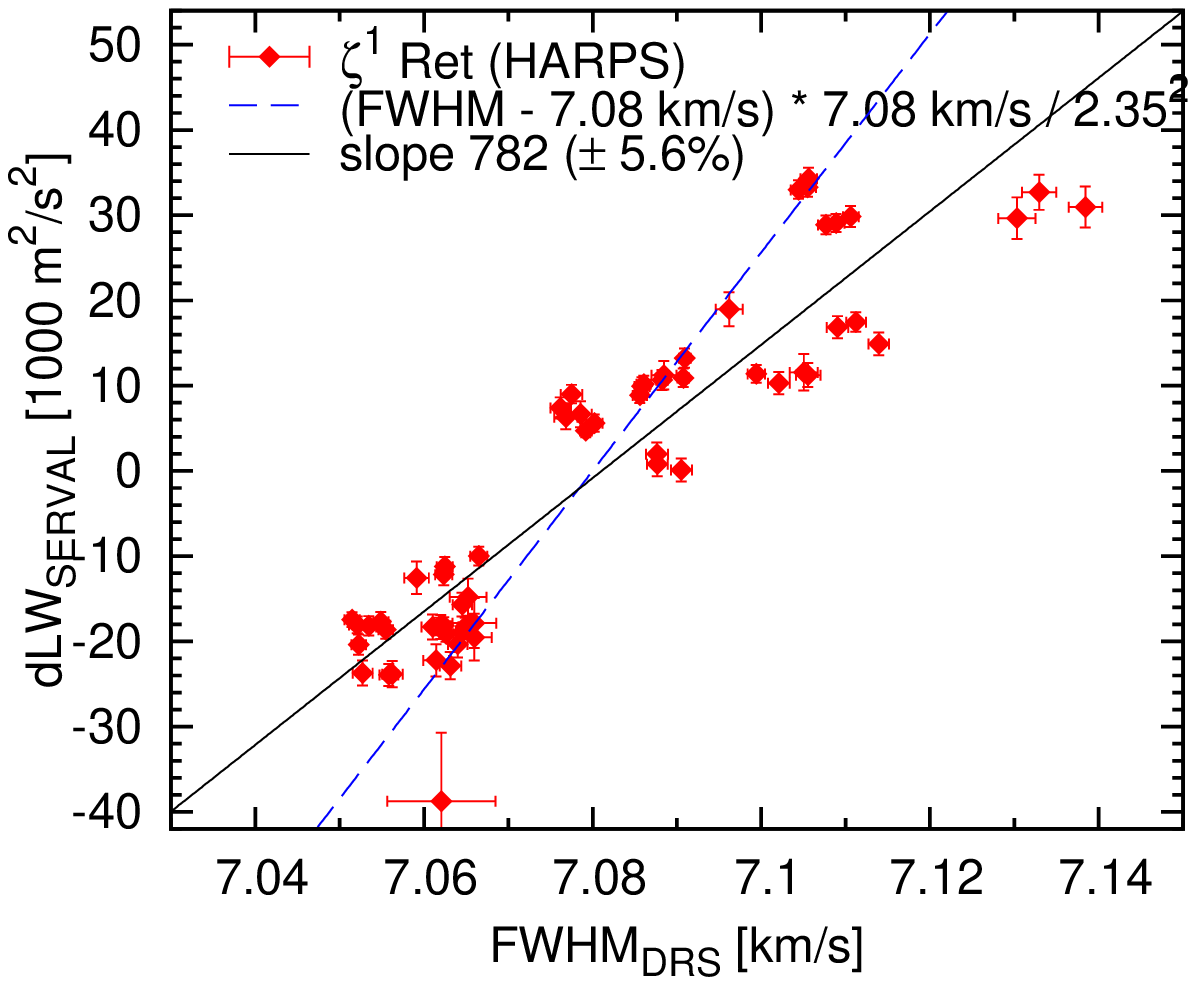}\includegraphics[width=0.5\linewidth]{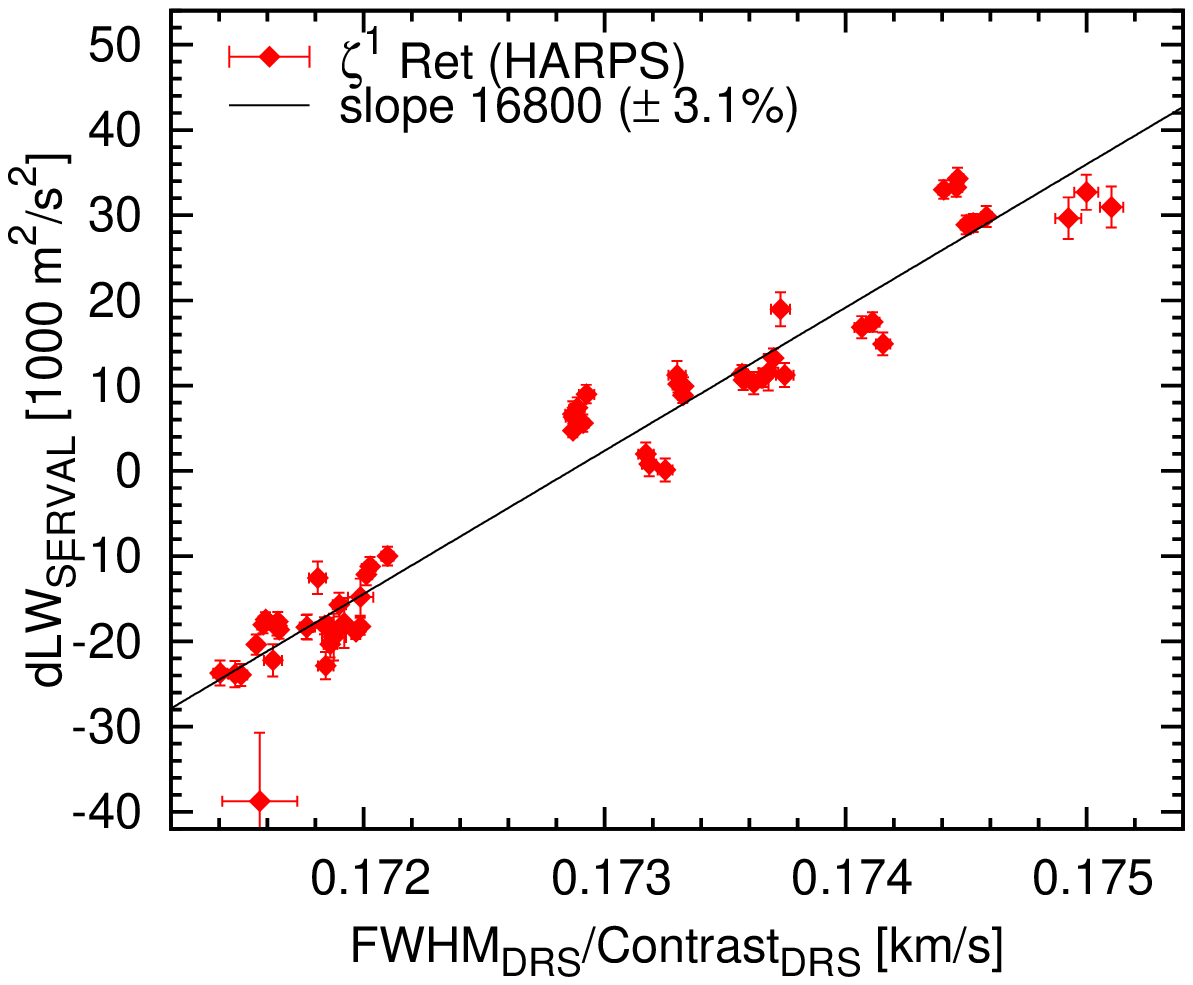}
\par\end{centering}

\caption{\label{fig:FWHM-dLW} Left: Correlation between the DRS-FWHM and the
SERVAL-dLW with HARPS data for the stars GJ~699 (top) and $\zeta^{1}$~Ret
(bottom). The solid black line indicates the best fitting linear trend;
the blue dashed line represents the theoretical prediction. Right: Same as
left panels, but the FWHM is divided by the contrast.}
\end{figure}

\paragraph{GJ~699} The comparison between dLW and FWHM, drawn
on the left and right axes, respectively, is shown
as a time series in the middle panel of Fig.~\ref{fig:GJ699}. Both indicators have relatively
similar error bars and vary significantly; we adopted $\epsilon_{FWHM}=2.35\epsilon_{RV}$. Using the relation $FWHM=2.35\sigma$ for
a Gaussian, we would expect $\Delta FWHM=\frac{FWHM\cdot\Delta FWHM}{FWHM}=\frac{2.35^{2}}{FWHM}\cdot\sigma\Delta\sigma$.
This theoretical prediction is shown as a blue line in Fig.~\ref{fig:FWHM-dLW}
(top left panel). We see that dLW values scatter around this prediction,
but the large uncertainty for the best fitting slope shows that the
dLW and FWHM correlate poorly for GJ~699. Surprisingly, we find a
much better correlation of dLW with FWHM divided by the contrast indicator,
which is the relative amplitude of the Gaussian fit (Fig.~\ref{fig:FWHM-dLW},
top right). This is currently not well understood, but it likely reminds
us that dLW and FWHM are not fully tracking the same effect. Indeed,
the dLW indicator assumes that the area of the Gaussian-shaped line,
i.e. the product of FWHM and contrast, remains constant, while a Gaussian
fit to the CCF decomposes both parameters simultaneously. Hence, we
could argue that additional contrast variations (on top of contrast
$\propto FWHM^{-1}$) could lead to biased dLW measurements. Those
contrast variations could be induced by imperfect background subtraction
or have stellar origin, albeit we believe the latter is less likely
for this quiet star. Alternatively, the poor dLW-FWHM correlation
for GJ699 could point to a bias when fitting a Gaussian function to
the CCF of an M dwarf, which is known to exhibit prominent side lobes
(see e.g. Fig.~7 in \citealp{Berdinas2017}). Neither dLW nor FWHM
correlate with RVs.

\begin{figure}
\begin{centering}
\includegraphics[width=1\linewidth]{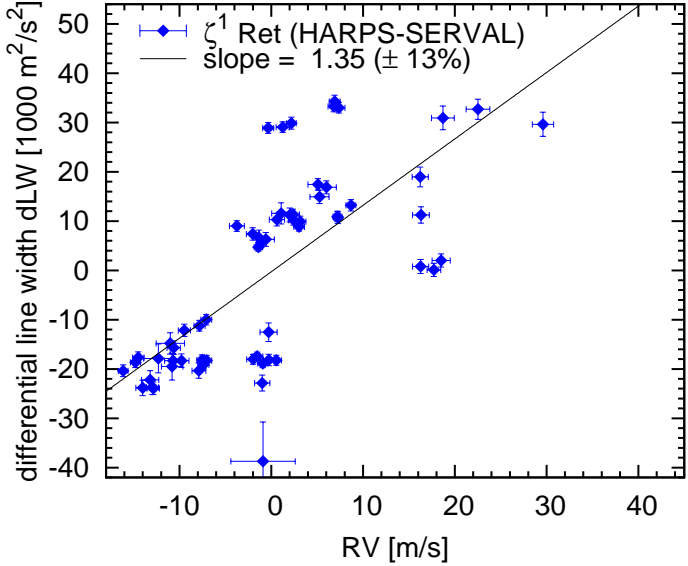}
\par\end{centering}

\caption{\label{fig:Zet1Ret-RV-dlw}Correlation between RV and dLW for HARPS
data of $\zeta^{1}$~Ret.}
\end{figure}

\paragraph{$\zeta^{1}$~Ret} The time series of dLW and FWHM already
indicates a similar behaviour (Fig.~\ref{fig:Zet1Ret}, middle).
Both, FWHM and dLW, correlate with RV (Fig.~\ref{fig:Zet1Ret-RV-dlw}).
There is also a direct correlation between dLW and FWHM; still the
correlation improves again when dividing the FWHM by the contrast
(Fig.~\ref{fig:FWHM-dLW}, bottom panels).

\paragraph{YZ~CMi} The dLW (Fig.~\ref{fig:YZCMi}, middle) reveals
a 2.776\,d period that we have already found in the RVs. The correlation
between dLW and RV is not linear (Fig.~\ref{fig:YZCMi-RV-csl}, bottom
and top), but we find that our observations follow a well-defined
path when we colour code the rotational phase with this period. The one
dLW outlier is an observation with an H$\alpha$ flare. The spectrum
has the highest H$\alpha$ core emission and noticable rising H$\alpha$
wings compared to the other spectra. Likely, the flare adds continuum
flux over the full spectrum, leading primarily to line contrast variations.
The flare event is not noticable in RV and chromatic index, although
the point has maximum RV and minimum chromatic index.

\subsection{\label{sub:YZ-CMi}Criterion 3: chromatic RV Index}

\paragraph{GJ~699} The chromatic index (Fig.~\ref{fig:GJ699},
right) does not vary significantly. This is expected for an inactive
and slowly rotating star.

\paragraph{$\zeta^{1}$~Ret} The chromatic index (Fig.~\ref{fig:Zet1Ret})
shows long-term variations that seem to be anti-correlated with the
RV long-term variations, which could be an activity cycle. However,
the chromatic index does not follow the short-term variations of RV
and dLW. Therefore, the chromatic index apparently has a complementary
role to line width indicators.

\begin{figure}
\begin{centering}
\includegraphics[width=1\linewidth]{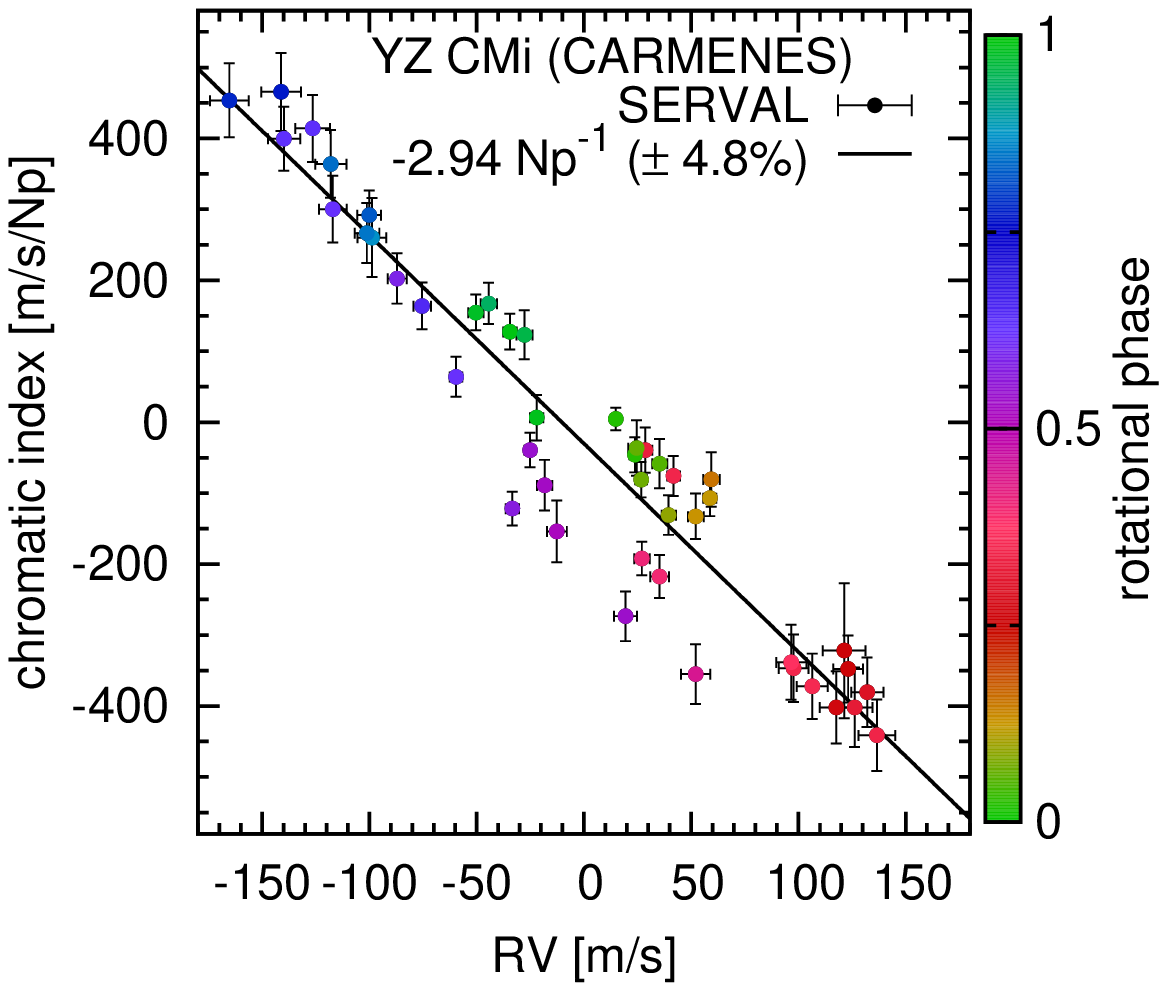}
\par\end{centering}

\begin{centering}
\includegraphics[width=1\linewidth]{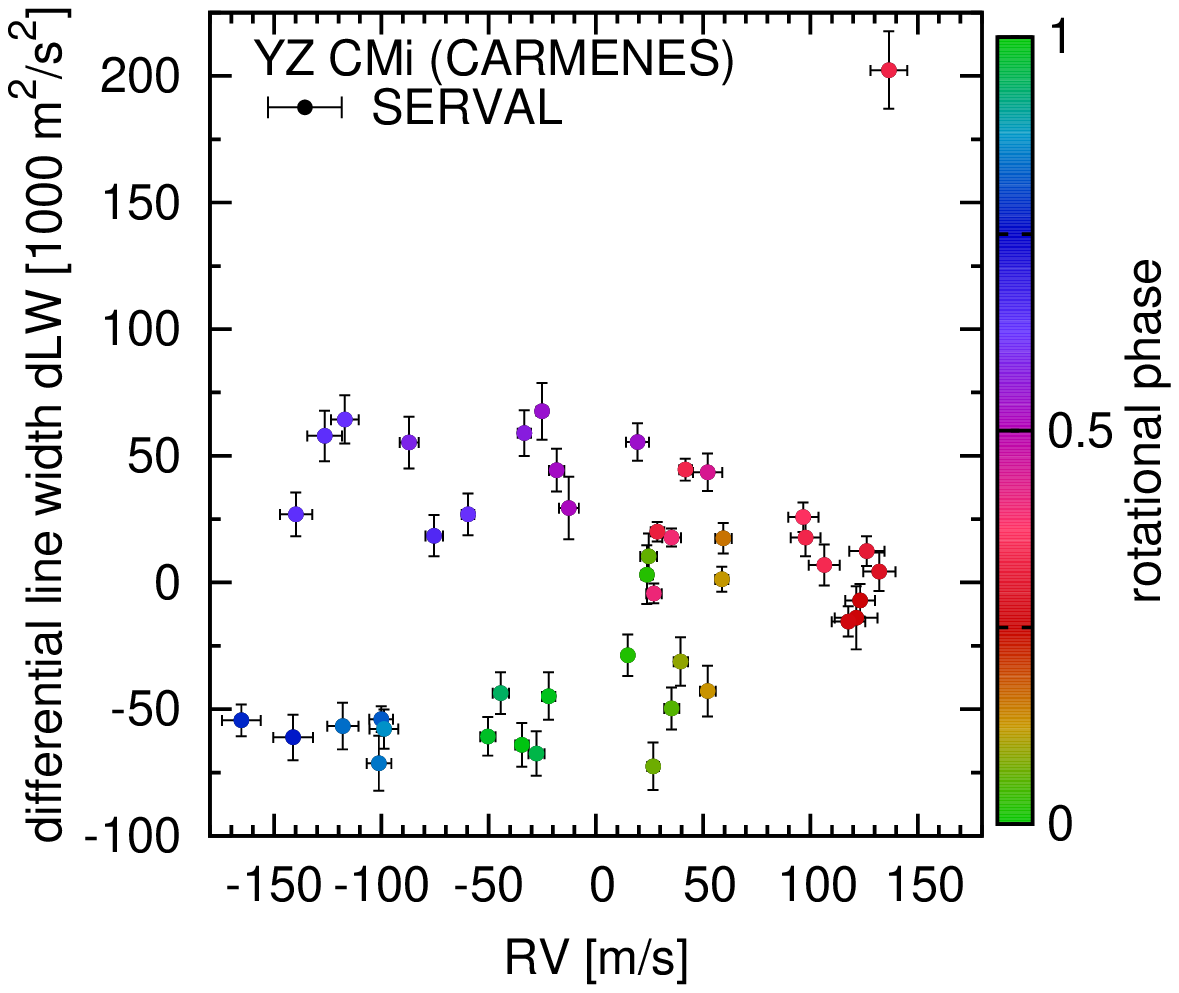}
\par\end{centering}

\caption{\label{fig:YZCMi-RV-csl}Top: Chromaticity (correlation between RV
and chromatic index) for 45 CARMENES VIS spectra of YZ~CMi. The left-most
(RV=165.3\,m/s, $\beta=453$\,m/s/Np) and right-most data points
(RV=136.6\,m/s, $\beta=-441$\,m/s/Np) are the values derived in
Fig.~\ref{fig:chromatic-index}. Bottom: RV-dLW correlation is shown.}
\end{figure}

\paragraph{YZ~CMi} In the top panel of Fig.~\ref{fig:YZCMi-RV-csl},
we show the chromatic index as a function of RV for our CARMENES observations
of YZ~CMi; both parameters are clearly anti-correlated. We call this
correlation and the corresponding slope \emph{chromaticity~}$\kappa$.
RV variations of $\pm150$\,m/s correspond to variations of the chromatic
index of $\mp450$\,m/s/Np. Of course, due to the anti-correlation,
the periodograms for chromatic index and RV have a similar shape and
the 2.776\,d period is also found in the chromatic index. For this
period, we colour code our data in Fig.~\ref{fig:YZCMi-RV-csl} according
to their rotational phase. One can clearly follow how the individual
observations are distributed throughout the rotational phase of YZ~CMi.

The chromatic index-RV correlation demonstrates that the dominating
effect in the RV variations of YZ~CMi is wavelength dependent, which
is consistent with the hypothesis that they are caused by co-rotating
features such as active regions on the surface of the star. The wavelength-dependent
temperature contrast between quiet and active regions is expected
to cause a variation of the RVs as a function of wavelength (e.g.
Fig.~12 in \citealp{Reiners2010}). Active regions (or spots) that
are somewhat cooler (or hotter) than the photosphere show less
contrast to the photosphere at longer wavelengths. The chromatic index
attempts to capture this dependence on wavelength. The advantage of
the chromatic index over chromospheric emission (Ca\textsc{~ii}~H\&K,
H$\alpha$) is that it comes with a directional sense, i.e. it can
be positive and negative. The chromatic index is directly connected
to RV because it is calculated from the same line profile deformation.

A negative chromaticity, as in case of YZ~CMi, means that the RV
scatter decreases towards redder wavelength. This amplitude decrease
can been seen in Fig.~\ref{fig:chromatic-index} and is predicted
with spot simulations \citep{Desort2007,Reiners2010}. The slope in
the correlation plot (Fig.~\ref{fig:YZCMi-RV-csl}, top) provides information
about spot temperatures. Moreover, we identify a separation between
the values of the chromatic index at rotational phases 0.5 and 1,
which would not be the case if the relation was simply linear and
is probably related to convective blue shift. A deeper interpretation
of these effects requires more detailed observations and goes beyond
the scope of this paper.

\subsection{Chromatic index with CARMENES and HARPS}

Using the example of YZ~CMi, we demonstrated that the chromatic
index in the wavelength range of CARMENES VIS (550--970\,nm) is a
powerful tool for understanding RV variations induced by stellar activity.
A closer look at the individual RVs per wavelength (Fig.~\ref{fig:chromatic-index})
shows that a fairly tight relation between RV and wavelength exists
for the wavelength range 600--920\,nm, but the five spectral orders
short of $\lambda=590$\,nm fail to follow this trend. We observe
this behaviour in the other exposures of YZ~CMi, too. The significance
of this effect for YZ~CMi and other stars needs more detailed investigation.
There are two plausible explanations for the lack of correlation at shorter wavelengths: first,
 the contrast grows too high in that active regions no longer contribute
to the observed spectrum in a significant way; and second that the intensity
of spectral features at short wavelengths also differs dra\-ma\-ti\-cal\-ly
between active and quiet regions \citep{Reiners2010}.

A possible way to investigate whether the chromatic index effect continues
down to shorter wavelengths is to determine RVs for individual spectral
orders of spectra taken with HARPS. This was attempted for the active
M4.5 dwarf AD~Leo by \citet[Fig.~9]{Reiners2013}. Interestingly,
the activity-induced RV amplitude reported there \emph{grows} with
wavelength (positive chromaticity $\kappa$), but the effect is more
than an order of magnitude weaker than what we observe in the CARMENES
data of YZ~CMi. Additionally, the two RV amplitudes shown for AD~Leo
by \citet{Reiners2013} that overlap with the CARMENES wavelength
range in fact show a much steeper wavelength dependence with the same
negative sign as observed in the CARMENES data for YZ~CMi. The comparison
between AD~Leo and YZ~CMi, however, remains inconclusive because
we might just be seeing very different effects in different stars.

In order to make a more useful comparison of the HARPS and CARMENES
wavelength ranges, we looked for targets that were observed with both
instruments, albeit not simultaneously. One of the few targets available
for this exercise is GJ~3379. Radial velocities and line indicators are shown in
Fig.~\ref{fig:GJ3379}. The datasets were taken at very different
times. The RVs from HARPS spectra exhibit an rms of 74\,m/s while
the CARMENES RVs show an rms of 21\,m/s. The intrinsic uncertainties
of the individual RV measurements are of the order of a few m/s in
both cases and negligible for this comparison.

\begin{figure}
\begin{centering}
\includegraphics[width=0.5\linewidth]{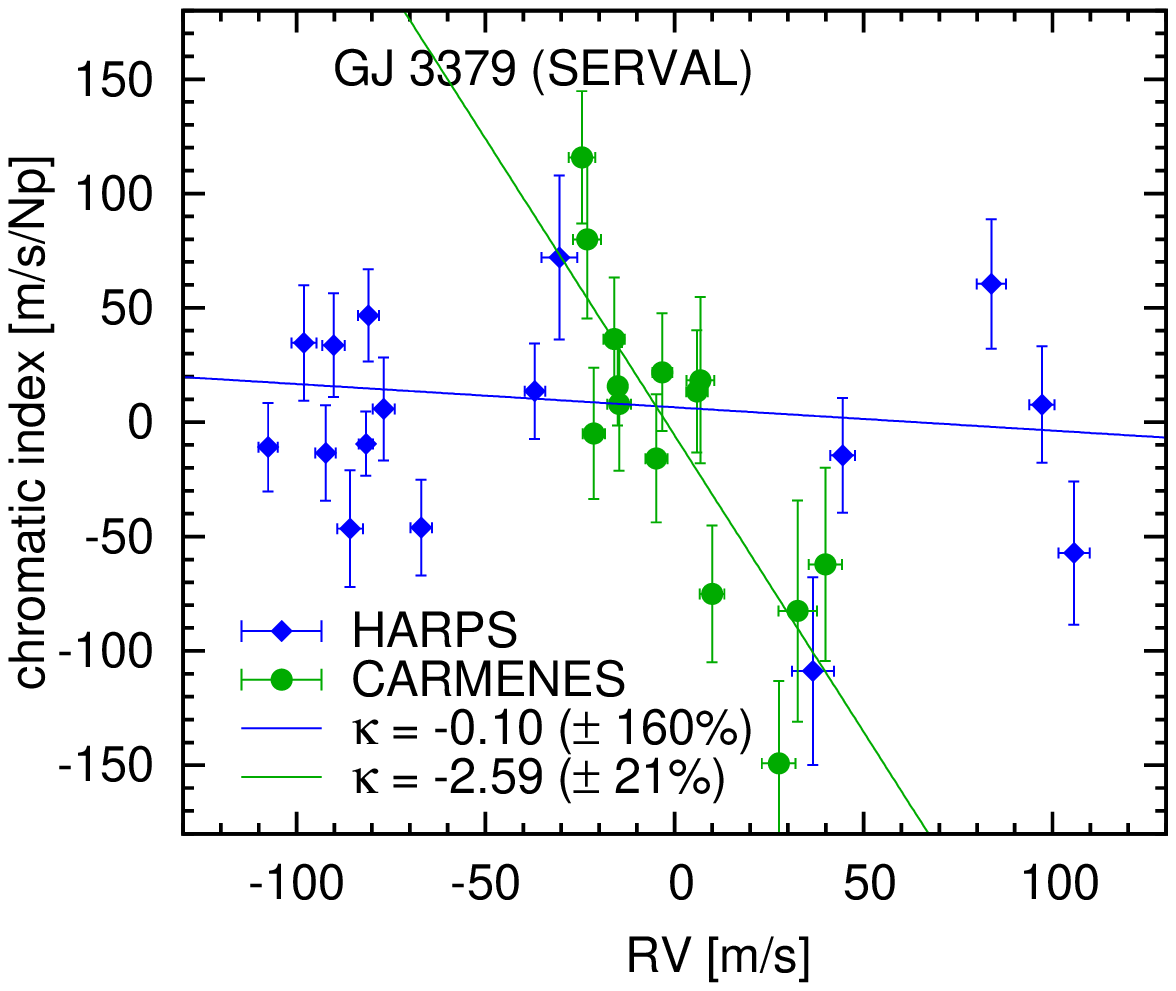}\includegraphics[width=0.5\linewidth]{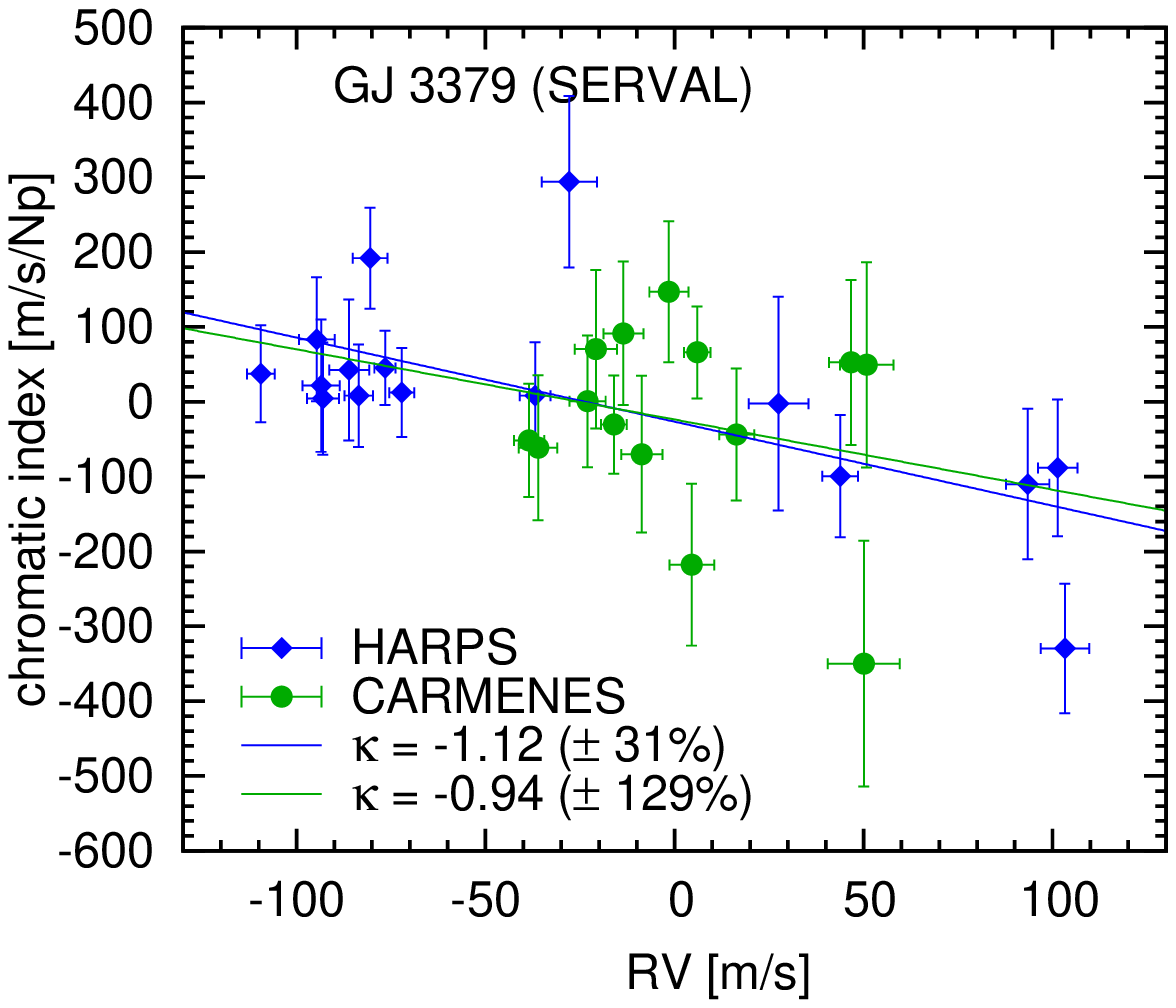}
\par\end{centering}

\caption{\label{fig:GJ3379 RV-csl}Radial velocity-chromatic index correlation for GJ~3379.
The fitted slope is the chromaticity of GJ~3379 in HARPS data (blue
diamonds) and CARMENES data (green circles) using their full spectral
range (left) and only the overlapping spectral region (right).}
\end{figure}

We calculated the chromatic index in both datasets and show the correlations
between chromatic index and RV in Fig.~\ref{fig:GJ3379 RV-csl}.
For our CARMENES data, chromatic index and RVs are correlated. The
relation can be described by a linear fit with a gradient of $-2.6\,\mathrm{Np}^{-1}$($\pm21\%$).
In the HARPS data, we find no correlation between chromatic index
and RV, although the scatter in RV is higher than in the CARMENES
data.

This comparison seems to indicate that the RVs of GJ~3379 show larger
scatter at bluer wavelengths and that the RV variations at longer
wavelengths are correlated with the chromatic index, while those at
bluer wavelengths are not. However, an alternative explanation is
that the RV jitter of the star changed during the course of a potential
activity cycle; the HARPS and CARMENES data are separated in time
by four years. One way to test this scenario is to compare RVs from
the same wavelength range in both datasets; HARPS and CARMENES overlap
in the range 550--690\,nm. Unfortunately, this wavelength range is
too short to draw definite conclusions about the chromatic
index, and we have already shown above that the slope is most useful
at longer wavelengths. Nevertheless, we find a correlation slope between
chromatic index and RV of about $-1\,\mathrm{Np}^{-1}$. Although
the scatter around this relation is relatively high, HARPS and CARMENES
data points are consistent with each other. The correlation
between chromatic index and RVs is much weaker than for the full CARMENES
wavelength range and the correlation itself is marginally significant.
On the other hand, the RVs calculated from the limited wavelength
range are precise enough to compare the two time series; in this case
we find that the rms of CARMENES RVs (30 m/s) is significantly lower
than the HARPS rms (77 m/s). It is unlikely that this effect is caused
by differences between the two instruments because for both we estimate
the internal uncertainties to be much smaller and comparable with
each other (4 m/s). Thus, for now the case remains undecided. We find
that GJ~3379 shows reduced activity jitter during the epoch observed
with CARMENES, while the RV rms was high when it was observed with
HARPS. Nevertheless, the CARMENES RVs show a correlation with chromatic
index when the full CARMENES wavelength range is taken into account.
We expect that such a correlation would still be visible when GJ~3379
exhibits larger RV variations, but simultaneous observations of the
HARPS and CARMENES wavelength ranges are required to settle this question.

\section{Summary}

We have presented our concept to derive high-precision RVs and have provided
details on the creation of a template. Following an iterative and
sequential approach we first derive approximate RVs measured against
an observed spectrum and then improve the template by co-adding all
observed spectra and recompute the RVs. The spectra are ``coadded''
via a weighted least-squares regression with a cubic $B$-spline.

The SERVAL code yields an output consisting of time series for high-precision
RVs and a number of spectral indicators useful for further diagnostics
as well as the high signal-to-noise templates partly cleaned from
telluric contamination taking advantage of barycentric shifts. We
find a similar performance when comparing the RV results from SERVAL
and the HARPS-DRS pipeline using data for GJ~699 and $\zeta^{1}$~Ret.
An additional example can also be found in \citet{Zechmeister2014},
where an older version of SERVAL was applied to \textasciitilde{}150
$\tau$~Cet spectra obtained during one night. This demonstrates
that SERVAL produces RVs at a 1\,m/s precision level.

Furthermore, we motivated a definition to study differential
changes in the spectral line widths. We scale the second derivative
of the template to the residuals from the fit for the best RVs and
have shown that it can be useful in practice. This differential indicator
nicely fits in our self-consistent framework of least-squares fitting
(for differential RVs) without the need for external templates. Of
course, this comes at the price that the differential results have
no external accuracy, i.e. the line width remains unknown. In contrast,
unbiased and accurate line width measurements for M dwarf spectra
with their overall blended lines can be obtained with least-squares
deconvolution \citep{Barnes2012MNRAS.424..591B}, but those are more
complicated algorithms that require accurate line positions as input.

The chromatic index, while very simple in its definition, turns out
to be a very powerful indicator to identify stellar activity in RV
signals. We have shown that it correlates very well with RVs for the
active M dwarf YZ~CMi. Hence it will help us to identify signals
that are associated with the intrinsic stellar variations rather than variations induced by Keplerian motion, to lower the detection
limit for active and inactive stars because it can be used to remove
and disentangle the activity contribution from the RVs, and to better
understand the physics of the stars, such as spot temperature or convective
blue shift. The chromatic index summarises in one value a first-order
effect of wavelength dependence in the RVs and is well suited for
an efficient analysis. Still, a more detailed investigation in smaller
wavelength ranges might further our understanding of the wavelength
dependence of stellar radial velocities \citep{Reiners2013,Feng2017}.
A large wavelength range, such as that offered by CARMENES, is essential
for this.
\begin{acknowledgements}
We thank J. Zhao, Y. Thiele, E.~Nagel, L. Nortmann, and G. Anglada-Escud\'{e}
for software testing and helpful discussions. M. Z. acknowledges support
from the Deutsche Forschungsgemeinschaft under DFG RE 1664/12-1 and
Research Unit FOR2544, project no. RE 1664/14-1. I. R. acknowledges
support by the Spanish Ministry of Economy and Competitiveness (MINECO)
and the Fondo Europeo de Desarrollo Regional (FEDER) through grant
ESP2016-80435-C2-1-R, as well as the support of the Generalitat de
Catalunya/CERCA programme. V. Wolthoff acknowledges funding from the
DFG Research Unit FOR2544, project no. RE 2694/4-1. VJSB are supported
by grant AYA2015-69350-C3-2-P from the Spanish Ministry of Economy
and Competiveness (MINECO). The UCM, CAB, and IAA team members acknowledges
support by the Spanish Ministry of Economy and Competitiveness (MINECO)
from projects AYA2016-79425- C3-1,2,3-P. CARMENES is an instrument
for the Centro Astron\'omico Hispano-Alem\'an de Calar Alto (CAHA,
Almer\'{\i}a, Spain). CARMENES is funded by the German Max-Planck-Gesellschaft
(MPG), the Spanish Consejo Superior de Investigaciones Cient\'{\i}ficas
(CSIC), the European Union through FEDER/ERF funds, and the members
of the CARMENES Consortium (Max-Planck-Institut f\"ur Astronomie,
Instituto de Astrof\'{\i}sica de Andaluc\'{\i}a, Landessternwarte
K\"onigstuhl, Institut de Ci\`encies de l'Espai, Insitut f\"ur
Astrophysik G\"ottingen, Universidad Complutense de Madrid, Th\"uringer
Landessternwarte Tautenburg, Instituto de Astrof\'{\i}sica de Canarias,
Hamburger Sternwarte, Centro de Astrobiolog\'{\i}a and Centro Astron\'omico
Hispano-Alem\'an), with additional contributions by the Spanish Ministry
of Economy, the German Science Foundation (DFG) through the Major
Research Instrumentation Programme and DFG Research Unit FOR2544 \textquotedblleft{}Blue
Planets around Red Stars\textquotedblright{}, the Klaus Tschira Stiftung,
the states of Baden-W\"urttemberg and Niedersachsen, and by the Junta
de Andaluc\'{\i}a.
\end{acknowledgements}
\bibliographystyle{aa}
\bibliography{serval}

\appendix

\section{\label{sec:Parabolic-interpolation}Parabolic interpolation}

We want to derive the interpolating parabola $y=ax^{2}+bx+c$ through
the three points $(x_{i},y_{i})$, $(x_{i-1},y_{i-1})$, and $(x_{i+1},y_{i+1})$.
The points should be equidistant, i.e. $x_{i+1}-x_{i}=\Delta x$ and
$x_{i}-x_{i-1}=\Delta x$. Now we centre points to $(x_{i},y_{i})$
and the three transformed points are $(0,0)$, $(-\Delta x,Y_{-1})$
and $(\Delta x,Y{}_{+1})$. Inserting into the parabola equation leads
to a system of three equations
\begin{align}
0 & =c\nonumber \\
Y_{-1} & =a\Delta x^{2}-b\Delta x\\
Y_{1} & =a\Delta x^{2}+b\Delta x\,.\nonumber 
\end{align}
The solution is
\begin{align}
a & =\frac{Y_{-1}+Y_{1}}{2\Delta x^{2}}=\frac{y_{i-1}-2y_{i}+y_{i+1}}{2\Delta x^{2}}\nonumber \\
b & =\frac{Y_{1}-Y_{-1}}{2\Delta x}=\frac{y_{i+1}-y_{i-1}}{2\Delta x}\\
c & =0\,.\nonumber 
\end{align}
Reforming the parabola equation to $y=y_{i}+a(x-x_{i}+\frac{b}{2a})^{2}-(\frac{b}{2a})^{2}+c$,
the extremum of the parabola is at
\begin{equation}
x_{c}=x_{i}-\frac{b}{2a}=x_{i}-\frac{\Delta x}{2}\cdot\frac{y_{i+1}-y_{i-1}}{y_{i-1}-2y_{i}+y_{i+1}}\label{eq:parabola_minimum}
\end{equation}
and the second derivative at $x_{c}$ is
\begin{equation}
y''=2a=\frac{y_{i-1}-2y_{i}+y_{i+1}}{\Delta x^{2}}\,,\label{eq:parabola_curvature}
\end{equation}
 which is the definition of the finite second derivative.

\section{\label{sec:Derivative-of-Gaussian}Derivatives of the Gaussian function}

We start with a Gaussian function with unit amplitude. Their first
and second derivatives (with respect to $x$) are given by
\begin{align}
f(x) & =\exp\left(-\frac{1}{2}\frac{x^{2}}{\sigma^{2}}\right)\label{eq:gauss_unitamp}\\
f'(x) & =-\frac{x}{\sigma^{2}}\exp\left(-\frac{1}{2}\frac{x^{2}}{\sigma^{2}}\right)=-\frac{x}{\sigma^{2}}f(x)\\
f''(x) & =\frac{1}{\sigma^{2}}\left(\frac{x^{2}}{\sigma^{2}}-1\right)f(x).
\end{align}

We seek to understand how the residuals should appear when we change the width while
keeping the amplitude. The derivative of Eq.~(\ref{eq:gauss_unitamp})
with respect to parameter $\sigma$ can be written as
\begin{align}
\frac{\partial f(x)}{\partial\sigma} & =\frac{x^{2}}{\sigma^{3}}\exp\left(-\frac{1}{2}\frac{x^{2}}{\sigma^{2}}\right)=\frac{x^{2}}{\sigma^{3}}f(x)\nonumber \\
 & =\sigma\frac{f'(x)f'(x)}{f(x)}\,.
\end{align}

When we are interested in the residuals while conserving the area
under the curves (Fig.~\ref{fig:dLW}, right panel), we have to include
the normalising factor $\frac{1}{\sigma}$
\begin{align}
g(x) & =\frac{1}{\sqrt{2\pi}\sigma}\exp\left(-\frac{1}{2}\frac{x^{2}}{\sigma^{2}}\right)\,.
\end{align}
The relation of $g'(x)$ and $g''(x)$ to $g(x)$ is the same as for
$f(x)$. However, the derivative with respect to\ $\sigma$ is
\begin{align}
\frac{\partial g(x)}{\partial\sigma} & =\frac{1}{\sigma}\left(\frac{x^{2}}{\sigma^{2}}-1\right)g(x)\nonumber \\
 & =\sigma g''(x)\,.\label{eq:dg_ds}
\end{align}

This suggests that we correlate the second derivative of the spectrum with
the residuals when we want to infer differential variations in the
width.

The mean value of the second derivative is
\begin{equation}
\int_{-\infty}^{\infty}g''(x)\mathrm{d}x=\left.g'(x)\right|_{-\infty}^{\infty}=0\,.
\end{equation}

\end{document}